\documentclass[journal]{IEEEtran}
\usepackage{amsmath,amsfonts}
\usepackage{algorithmic}
\usepackage{algorithm}
\usepackage{array}
\usepackage[caption=false,font=normalsize,labelfont=sf,textfont=sf]{subfig}
\usepackage[colorlinks=false,linkcolor=blue,citecolor=red]{hyperref}
\usepackage{textcomp}
\usepackage{stfloats}
\usepackage{url, xcolor}
\usepackage{verbatim}
\usepackage{graphicx}
\usepackage{cleveref}
\usepackage{cite}
\usepackage{booktabs} 
\usepackage{multirow} 
\usepackage{tabularx}

\begin{document}

\title{A Survey for Deep Reinforcement Learning Based Network Intrusion Detection}

\author{Wanrong Yang{\href{https://orcid.org/0000-0003-0216-3762}{\includegraphics[width=1em,height=1em]{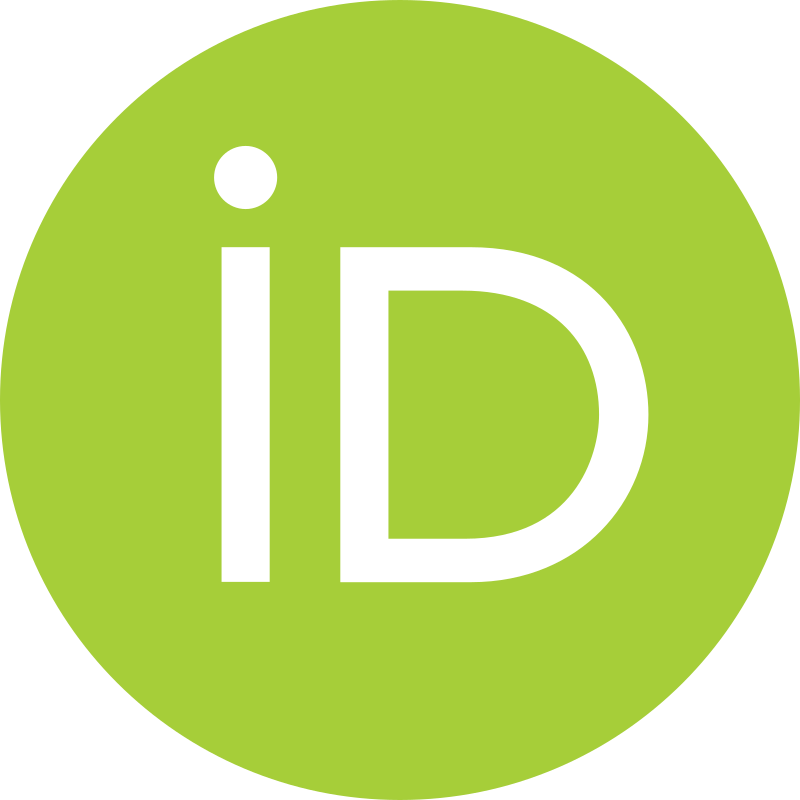}}},
        Alberto Acuto{\href{https://orcid.org/0000-0003-0753-5131}{\includegraphics[width=1em,height=1em]{orcid.png}}},
        Yihang Zhou{\href{https://orcid.org/0000-0001-6354-1259}{\includegraphics[width=1em,height=1em]{orcid.png}}},
        Dominik Wojtczak{\href{https://orcid.org/0000-0001-5560-0546}{\includegraphics[width=1em,height=1em]{orcid.png}}}
\thanks{This work was supported by the Engineering and Physical Sciences Research Council (EPSRC), through grants number EP/X017796/1 and EP/X03688X/1. We would like to express our sincere appreciation for research support from Centre for Doctoral Training (CDT) in Distributed Algorithm, University of Liverpool. Specifically, many thanks for all the help from Prof Simon Maskell, Kelli Cassidy, Elizabeth Gannon and Big hypotheses group. Thanks to Qingyuan Wu for kind and practical suggestions on the manuscript. In the end, we would like to say thanks to our industrial partners, Dr. Stephen Pasteris and Dr. Chris Hicks from Alan Turing Institute.}
\thanks{Wanrong Yang, Dominik Wojtczak are with the Department of Computer Sciences, University of Liverpool, Ashton Street, Liverpool, L69 3BX, United Kingdom. Dominik Wojtczak is also the head of the Cybersecurity Institute at the University of Liverpool. Email at \href{mailto:wanrong.Yang@liverpool.ac.uk}{wanrong.yang@liverpool.ac.uk} and \href{mailto:d.wojtczak@liverpool.ac.uk}{d.wojtczak@liverpool.ac.uk}}
\thanks{Alberto Acuto is with the Department of Electrical Engineering and Electronics, University of Liverpool, L69 3GJ, United Kingdom. Email at \href{mailto:a.acuto@liverpool.ac.uk}{a.acuto@liverpool.ac.uk}}
\thanks{Yihang Zhou is with the Shenzhen Institute of Advanced Technology, Chinese Academy of Sciences, Shenzhen 518055, P.R. China. Email at \href{mailto:yh.zhou2@siat.ac.cn}{yh.zhou2@siat.ac.cn}}
\thanks{Yihang Zhou and Dominik Wojtczak are co-corresponding authors.}
}

\maketitle

\begin{abstract}
Cyber-attacks are gradually becoming more sophisticated and highly frequent nowadays, and the significance of network intrusion detection systems has become more pronounced. This paper investigates the prospects and challenges of employing deep reinforcement learning technologies in network intrusion detection. It begins with an introduction to the fundamental theories and technological frameworks of deep reinforcement learning including classic deep Q-network and actor-critic algorithms, followed by a review of essential research that has leveraged deep reinforcement learning for network intrusion detection in recent years. This research assesses these challenges and efforts in terms of model training efficiency, the detection capabilities for minority and unknown class attacks, improved network feature selection and unbalanced dataset issues. Performances of deep reinforcement learning models are comprehensively investigated. The findings reveal that although deep reinforcement learning shows promise in network intrusion detection, many of the latest deep reinforcement learning technologies are yet to be fully explored. Some deep reinforcement learning based models can achieve state-of-the-art results in some public datasets, in some cases, even better than traditional deep learning methods. The paper concludes with recommendations for the enhanced deployment and testing of deep reinforcement learning technologies in real-world network scenarios to further improve their application. Special emphasis is placed on the Internet of Things intrusion detection. We offer discussions on recently proposed deep architectures, revealing possible future policy functions used for deep reinforcement learning based network intrusion detection. In the end, we propose integrating deep reinforcement learning and broader generative methods and models to assist and further improve their performance. These advancements aim to address the current gaps and facilitate more robust and adaptive network intrusion detection systems.
\end{abstract}


\section{Introduction}
Cybersecurity is a significant challenge in the age of global informatization \cite{ahsan2022cybersecurity}. As digital lifestyles become increasingly prevalent, our reliance on cybersecurity and the need for it is growing rapidly as well \cite{sarker2023machine}. Nowadays, cybercriminal activities are inflicting substantial economic losses on various industrial and government sectors. According to live data from the AV-TEST Institute, more than 450,000 new malicious programs are detected every day, with over 1.2 billion instances of malware emerging within 2023 alone, and these numbers are still on the rise \cite{geer2020market}. These malicious programs infiltrate personal and corporation digital systems through a wide range of vulnerabilities, posing serious threats to personal privacy, commercial secrets, and other sensitive records. It is estimated that cybersecurity-related issues have caused a total loss of up to 400 billion US dollars to the global economy \cite{fischer2014cybersecurity}. According to Cybersecurity Ventures, cybercrime has caused \$1.5 trillion in losses \cite{ventures20192019} in 2019, a figure that is expected to climb to \$9.5 trillion US dollars by 2024 \cite{ventures2024cybercrime}. Moreover, critical national infrastructures from energy, healthcare sectors, and port automation systems, are becoming primary targets for cyber-attackers \cite{ferrag2020deep}. According to 2024 World Economic Forum, 94\% of government leaders and business executives believe that their organizations are still at a lower defence level when facing cyber-attacks \cite{GlobalCybersecurityOutlook}. In short, the importance of cybersecurity is set to grow significantly worldwide.

Network Intrusion Detection (NID) is a crucial defence mechanism in the field of cybersecurity. It effectively protects computers and other digital devices from external attacks \cite{tsai2009intrusion}. It was first proposed in 1994 \cite{mukherjee1994network} and later described as integrating information extracted from computers to identify resource abuse within the network and attacks originating from outside entities \cite{caberera2000statistical}. Basically, intrusion detection systems (IDS) can be categorized into \emph{Network Intrusion Detection Systems} (NIDS), which is based on the observation of network traffic between different nodes \cite{thottan2003anomaly} and \emph{Host-based Intrusion Detection Systems} (HIDS), which means monitoring activities on a specific host, including applications being used and file systems being accessed \cite{newman2009computer}. The primary purpose of NID is to prevent network attacks by identifying abnormal traffic or access operations \cite{tsai2009intrusion}. Objectives of network attacks are becoming increasingly complex, traditional signature-based methods that identify known attacks based on pattern matching of known signatures have fallen behind anomaly-based detection approaches \cite{chou2021survey}. Since the anomaly-based detection has a higher efficiency and dynamic adaptability, it is now widely accepted by the NID community \cite{bhattacharya2021investigation}.

\begin{figure*}[ht]
    \centering
    \includegraphics[width=\textwidth]{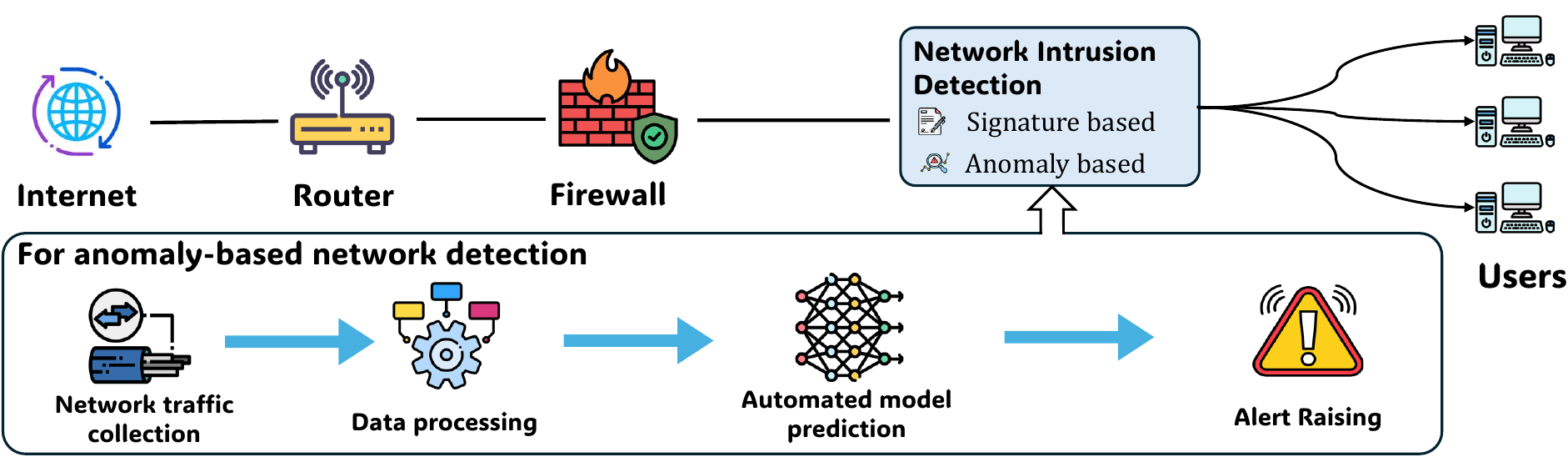}
    \caption{An overview of network intrusion detection and specific working flow of anomaly-based network intrusion detection.}
    \label{overview}
\end{figure*}

Artificial intelligence (AI) plays an essential role in NID \cite{ferrag2020deep, dixit2021deep}. \Cref{overview} shows an overview of network intrusion detection and specific working flow of anomaly-based network intrusion detection. The system is positioned between the Internet and internal users, with traffic flowing through a router and firewall before reaching the Network Intrusion Detection System, which employs both signature-based detection for known attack patterns and anomaly-based detection for novel threats. The anomaly-based detection pipeline, detailed in the lower section, consists of four sequential stages: network traffic collection from network devices, data processing for feature extraction and normalization, automated model prediction using neural network algorithms, and alert raising when potential intrusions are detected. This dual-detection approach combines traditional pattern-matching techniques with advanced machine learning methods to provide robust, real-time protection against both known and emerging network security threats. Traditional machine learning (ML) algorithms, including supervised and unsupervised learning e.g., Support Vector Machines (SVM), K-Nearest Neighbors (KNN), Random Forests (RF), and Multilayer Perceptron (MLP), have made improvements for NIDS in multiple ways \cite{ahmad2021network}. Further, deep learning (DL), by constructing deep neural networks, is capable of learning and fitting highly complex patterns and features from large amounts of given training data \cite{lopez2020application}. It can effectively learn and simulate patterns of normal behaviour in network traffic, thereby identifying abnormal activities or potential intrusions \cite{dixit2021deep, ahmad2021network, lopez2020application}. Compared with traditional ML methods, critical features in network intrusion could be extracted automatically by deep learning without as much laborious feature selection.

However, the DL model hugely relies on large, high-quality datasets \cite{li2019learning}. Keeping up to date with large-scale real-world cyber intrusion data is both time-consuming and labour-intensive. In addition, utilizing outdated datasets could potentially compromise the generalization capabilities of DL models \cite{damasevicius2020litnet}. Reinforcement Learning (RL), a subset of ML, imitates human learning strategies more closely than any other ML approach due to its ability to acquire knowledge from its own experiences by navigating and leveraging unfamiliar environments, and so is considered a potential solution to this problem \cite{nguyen2021deep}. Building on the principles of RL, Deep Reinforcement Learning (DRL) leverages neural networks to manage complex, high-dimensional input spaces. With outstanding decision-making and optimal control skills, DRL algorithms have achieved overwhelming success in many different fields, from real-world applications, e.g., drone racing \cite{kaufmann2023champion}, autonomous driving \cite{isele2018navigating}, biological data mining \cite{mahmud2018applications}, natural language processing \cite{keneshloo2019deep}, autonomous surgery \cite{nguyen2019new}, drug design \cite{popova2018deep} to virtual games domain, e.g., the game of Go \cite{silver2017mastering}, StarCraft II \cite{sun2018tstarbots}. Furthermore, because of the ability to dynamically adapt to the environment, DRL has been widely applied in cybersecurity, including in areas of NID \cite{mathew2021deep,sewak2023deep} and adversarial simulation enhancement \cite{oh2023applying}.

Some surveys focused on the general AI in intrusion detection \cite{sowmya2023comprehensive} or DRL in general cybersecurity \cite{nguyen2021deep}. However, there has been limited research on applying deep reinforcement learning to network intrusion detection. Thus, there is a need for a more detailed and comprehensive survey specifically focused on DRL and network intrusion detection. This paper primarily concentrates on the exploration of DRL applications within the domain of NID over the past five years. It aims to provide a systematic review of the most current advancements in RL applications for NID, endeavoring to elucidate how RL is revolutionizing and enhancing NID systems. To achieve this, firstly, preliminary knowledge of RL is presented including the Markov decision process, Q-learning, Deep Reinforcement Learning, Inverse Reinforcement Learning and how we evaluate the performance of RL-based NID model. Then, we retrospectively investigate representative works of DRL-based NID in the past few years focused on the overview of the most often used datasets, efforts on network feature engineering, handling unbalanced datasets, improving training efficiency and identifying minority intrusions. Finally, we present our general discussion based on these representative works hoping to provide some ideas and future research directions.

\begin{figure}[ht]
    \centering
    \includegraphics[width=0.9\linewidth]{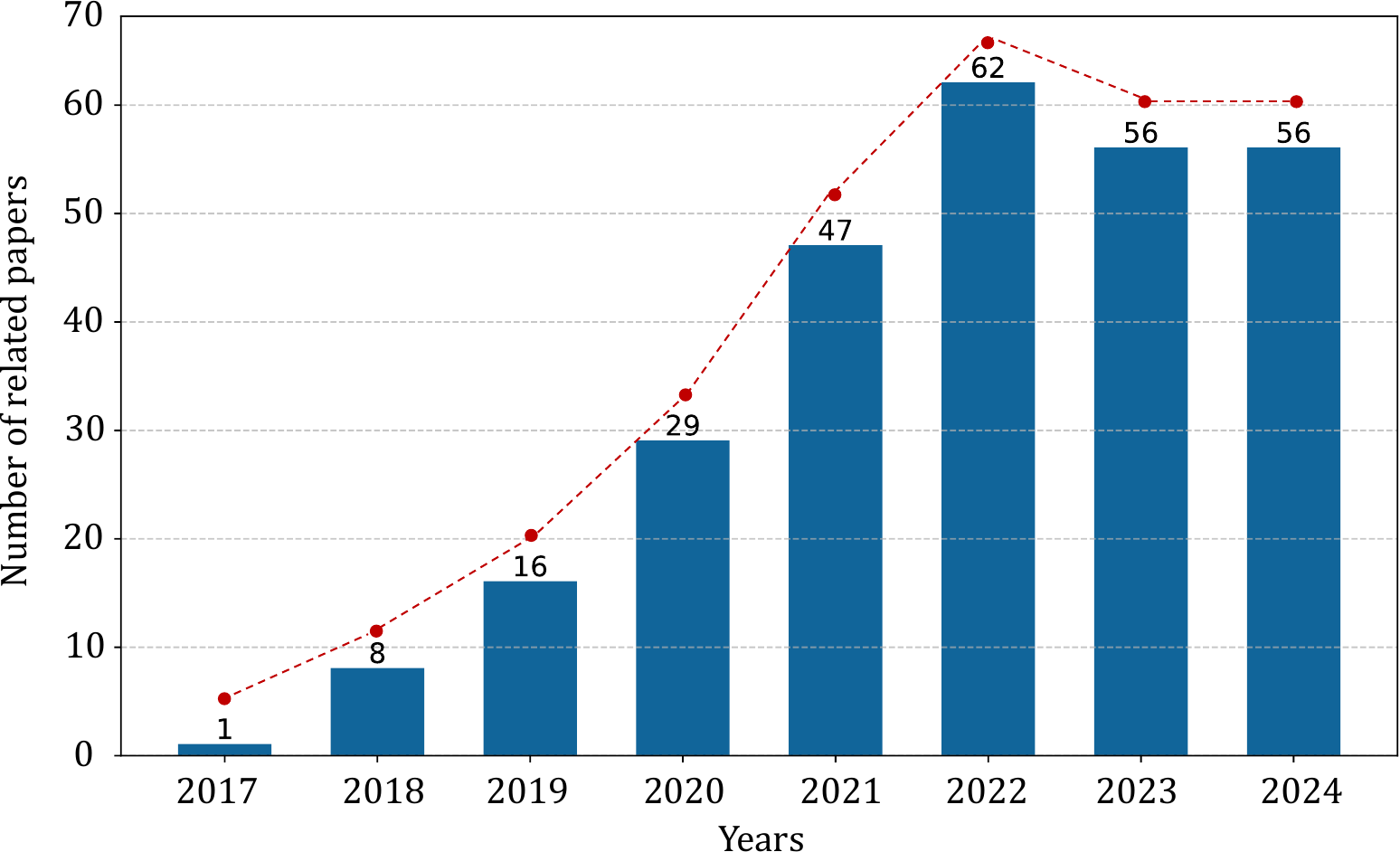} 
    \caption{Number of related publications since 2017, based on a Web of Science search using the keywords “reinforcement learning” and “network intrusion.”}
    \label{trend}
\end{figure}

\section{Reinforcement Learning Basics}
\subsection{Overview}
Reinforcement Learning is quite different from popular supervised and unsupervised learning in ML, which are typically driven by large example data. It involves a decision-making agent that starts without any prior knowledge and learns through its own experience. This is achieved by repeated and random interactions with an environment, allowing the agent to acquire essential knowledge to make an informed decision \cite{ladosz2022exploration}. 

A classic RL system is, in general, comprised of an environment, agent, policy, reward, and value function. \cite{ladosz2022exploration}. \textbf{Agent} means a decision-making, goal-seeking and highly interactive virtual entity. \textbf{Environment} refers to everything that the agent interacts with: applying action to it and receiving feedback from it. \textbf{Policy} $\pi$ determines which action the agent takes when in a given state of the environment. It could be a function or a simple lookup matrix. \textbf{Reward} is quantitative feedback from the environment after the agent takes one specific action followed by a state. The positive and negative reward describes how “good” or “bad” the action with regarding to the final goal of the agent. \textbf{Value function} is used to evaluate the quality level of a state or action. Thus, it is divided into state-value function and action-value function.

\subsection{Markov decision process}
\begin{figure}[ht]
    \centering
    \includegraphics[width=0.8\linewidth]{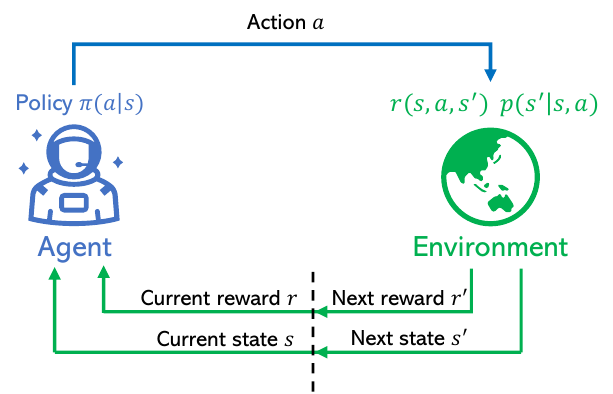} 
    \caption{Agent interaction with environment based on MDP in RL.}
    \label{fig:MDP}
\end{figure}

Markov decision process (MDP) is the principal framework for decision in stochastic and uncertain environment \cite{puterman1990markov}. It assumes that an agent can observe the current state $s_{t}$, and choose to take an action $a_{t}$. Then, the agent moves to the next state $s_{t + 1}$. Normally, it is described as a tuple $\left(S, A , P , R , \gamma\right)$ with the following five essential elements. $S$ is a state space, including all states that can be observed by the agent. $A$ is an action space, i.e., the set of all actions that can be taken by the agent. $p \left(s^{'}|s, a\right) \in P$ describes the probability of transferring to a specific state $s^{'}$ by taking action $a$ in the given state $s$. Reward function $R: S \times A \rightarrow \mathbb{R}$ gives the expected immediate reward $R(s,a) = \mathbb{E}[r_t | s_t=s, a_t=a]$ that the agent receives after taking action $a$ in state $s$. Alternatively, the reward can be defined as $r(s, a, s') \in \mathbb{R}$, representing the reward received when transitioning from state $s$ to $s'$ via action $a$, where $R(s,a) = \sum_{s' \in S} p(s'|s,a) r(s,a,s')$. $\gamma$ is a discount factor that can be used to indicate short-term or long-term importance of rewards. The fundamental property of an MDP is that the next state $s_{t + 1}$ depends only on the state $s_{t}$ and the action $a_t$ that the agent has taken. The interaction process between the agent and the environment in a Markov decision process is well illustrated in \Cref{fig:MDP}.

The final goal of the agent is to maximize long-term rewards. State-value function $V \left(\right. s \left.\right)$ can estimate the value of a state, which starts from a specific state $s$ following the action chosen by current policy $\pi$ of the agent, where $\pi$ is also a probability function. It can be utilized to approximate how much long-term reward the agent can gain in all next states based on the current $\pi$ shown in \Cref{state-value-function}. 

\begin{equation}
V(s) = \sum_{a \in A} \pi(a | s) \left( R(s, a) + \gamma \sum_{s' \in S} p(s' | s, a) V(s') \right)
\label{state-value-function}
\end{equation}
Firstly, the agent follows policy $\pi$ (decision-making policy) to take an action $a$ in state $s$ to gain an immediate reward $R(s,a)$. And over all possible next states $s'$, we have discounted sum of possible $V(s')$ (expected future returns) by adding discount factor $\gamma$ (future reward weight). Furthermore, \emph{Bellman optimality} \cite{rosu2002bellman} equations define optimal state-value function $V^*(s)$ to get the maximum return for each state and optimal action-value function $Q^*(s,a)$ to get the maximum return for each state-action pair, shown in \Cref{V_star} and \Cref{Q_star}.
\begin{equation}
V^*(s) = \max_a \left( R(s,a) + \gamma \sum_{s' \in S} p(s'|s,a)V^*(s') \right)
\label{V_star}
\end{equation}

\begin{equation}
Q^*(s,a) = R(s,a) + \gamma \sum_{s' \in S} p(s'|s,a) \max_{a'} Q^*(s',a')
\label{Q_star}
\end{equation}




\subsection{Q-learning}
Q-learning is a classic value-based algorithm in RL \cite{watkins1992q}. It allows the agent to learn how to select actions in a given state to maximize the expected total reward. \cite{clifton2020q}. The agent updates the value of action-state pair (Q value) by exploring possible actions of the state ($\epsilon$-greedy strategy \cite{singh2000convergence}). The update of Q value is based on the current Q value, immediate reward, and maximum Q value of next state. \Cref{Q_learning} is normally used to update the Q value. In the equation, the Q value $Q(s_t, a_t)$  represents the expected utility of taking action $a_t$ in the current state $s_t$, considering both the immediate reward $r_t$ and the expected rewards of future states.

\begin{equation}
Q(s_t, a_t) \leftarrow Q(s_t, a_t) + \alpha [r_t + \gamma \max_a Q(s_{t+1}, a) - Q(s_t, a_t)]
\label{Q_learning}
\end{equation}

Q-learning typically employs an $\epsilon$-greedy exploration strategy: with probability $1-\epsilon$, the agent selects the action $a_t = \arg\max_{a} Q(s_t, a)$ that maximizes the Q-value (exploitation); with probability $\epsilon$, it randomly selects an action from the action space (exploration). This balances exploration of the environment with exploitation of learned knowledge. $r_t$ is the immediate reward that the agent obtained by taking $a_t$. $Q(s_{t+1},a)$ means all possible Q value corresponding to all possible action $a$ in $s_{t+1}$. Discounted factor $\gamma$ is used to determine if the agent should focus on long-term reward, the smaller $\gamma$ is, the more short-sighted the agent becomes. Finally, the learning rate $\alpha \in (0,1]$ controls the magnitude of the Q-value update, determining the extent to which newly observed information overrides the existing estimate. A higher $\alpha$ places more weight on recent experience, while a lower $\alpha$ preserves more historical knowledge.

\subsection{Deep Reinforcement Learning}

\begin{figure}[ht]
    \centering
    \includegraphics[width=0.7\linewidth]{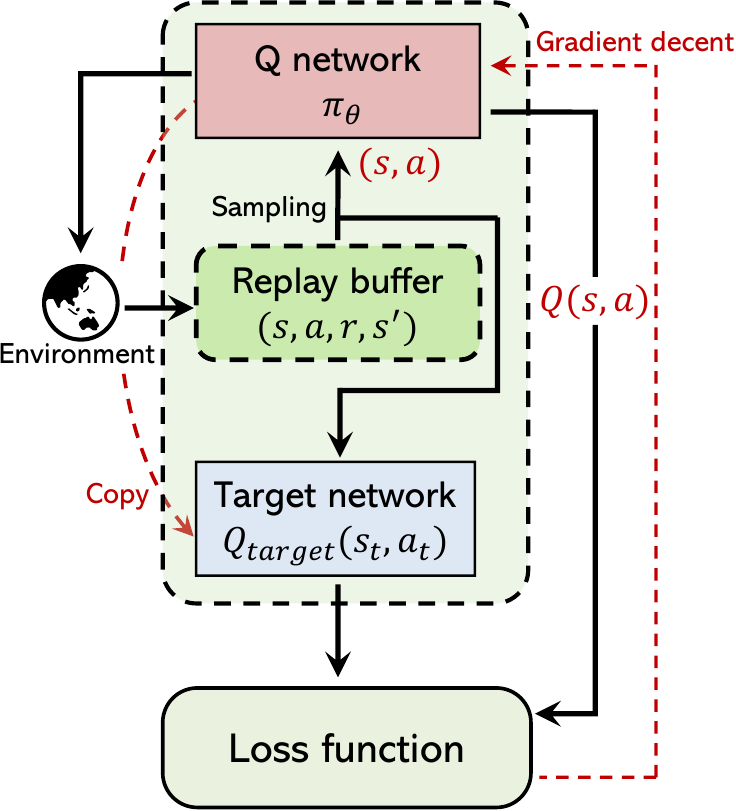} 
    \caption{Classic deep Q network process}
    \label{qlearning}
\end{figure}

\subsubsection{Deep Q Network, DQN}
Deep reinforcement learning (DRL), shown in \Cref{qlearning}, is a major progress in RL \cite{li2017deep}. It allows agents to handle high-dimensional inputs (e.g., images and audio), which pose challenges for typical Q-learning as they would require a massive table to store state-action pair values—typically unacceptable given memory constraints. DQN \cite{huang2020deep} is a classic DRL algorithm that leverages the dynamic transition of experience replay and utilization of a target network.

The experience replay buffer in DQN is a memory storage mechanism that retains past experiences in the form of state, action, reward, and next state tuples. It allows the agent to break the temporal correlations between consecutive experiences by randomly sampling from this buffer to train the Q-network. This process stabilizes and enhances the learning process by ensuring that updates are based on a diverse set of past experiences, rather than being dominated by recent, possibly highly correlated, events.

Specifically, the agent interacts with the environment using an $\epsilon$-greedy strategy (typically with high initial $\epsilon$ for exploration) and network parameters initialized as $\theta_0$ to collect initial experiences $(s_t, a_t, r_t, s_{t+1})$ that populate the experience replay buffer. Based on this initial knowledge, the target network (also initialized by $\theta_0$) could calculates the target Q value by using the Bellman equation for data in replay buffer. Then, in the training process, a minibatch of transitions is randomly sampled from the experience replay buffer to update the parameters $\theta$ of the Q-network by minimizing the loss function shown in \Cref{DQN_Loss}. The loss measures the mean squared error between the predicted Q-value $Q(s_t, a_t; \theta)$ and the target Q-value $r_t + \gamma \max_{a'} Q(s_{t+1}, a'; \theta^-)$, where $\theta^-$ denotes the parameters of the target network.

\begin{equation}
L(\theta) = \mathbb{E}[(r_t + \gamma \max_{a'} Q(s_{t+1}, a'; \theta^-) - Q(s_t, a_t; \theta))^2]
\label{DQN_Loss}
\end{equation}

After the Q-network parameters are updated to $\theta$, the agent continues to interact with the environment to collect new experiences, which are added to the experience replay buffer (oldest experiences may be discarded if the buffer is full). The target network parameters $\theta^-$ remain fixed for a predetermined number of training steps (e.g., every $C$ steps) and are then periodically synchronized with the current Q-network parameters: $\theta^- \leftarrow \theta$. This periodic update mechanism ensures that the target values remain stable during training, which is crucial for convergence. Alternatively, soft updates can be used: $\theta^- \leftarrow \tau \theta + (1-\tau)\theta^-$ where $\tau \ll 1$. The training will stop when the maximum accumulated reward has been achieved or the loss has converged to a stable range.

\subsubsection{Policy Gradient}
The policy gradient method \cite{silver2014deterministic} is employed to optimize decision-making policies in reinforcement learning directly, without relying on a value function. The principle, different from value function-based methods, is to optimize policy parameters directly and then to achieve the maximum total reward. Policy Gradient Theorem \cite{sutton1999policy} is the very foundation of policy gradient. Let a parametric policy $\pi(a|s, \theta)$, choosing action $a$ given a state $s$ by a probability $p = \pi_\theta$, then, the performance of the policy can be quantified as follows. $R(\tau)$ means return from a single trace. It refers to one complete trajectory or episode of sequential interactions between an agent and environment, consisting of a sequence of states, actions, and rewards from start to finish. A policy gradient can be described in \Cref{pg_1}, \Cref{pg_2} and \Cref{pg_3}:

\begin{equation}
J(\theta) = E_{\tau \sim \pi_\theta}[R(\tau)]
\label{pg_1}
\end{equation}

\begin{equation}
\nabla_\theta J(\theta) = \mathbb{E}_{\tau \sim \pi_\theta} \left[ \sum_{t=0}^{T-1} \nabla_\theta \log \pi_\theta(a_t|s_t) G_t \right]
\label{pg_2}
\end{equation}

\begin{equation}
G_t = \sum_{k=t}^\tau \gamma^{k-t} r_k
\label{pg_3}
\end{equation}

where $T$ denotes the time horizon of an episode (which may be finite or infinite depending on the problem setting), and $\tau = (s_0, a_0, r_0, s_1, a_1, r_1, ..., s_T)$ represents a complete trajectory, and $k$ represents a future time point starting from a specific initial time denoted by $t$. As time passes by, $k$ gradually increases. It is desired that the contribution of rewards received at farther future time points to the current state decreases gradually. This is achieved through the exponential term $\gamma^{k-t}$, where $\gamma \in [0,1]$ is a fixed discount factor that remains constant throughout training. The geometric decay $\gamma^{k-t}$ ensures that rewards further in the future have progressively smaller influence on the current state value. The initial trajectories are gathered from random interactions between the agent and the environment over a fixed duration. These trajectories are used only once. After all trajectories in a set have been utilized, the agent restarts interactions with the environment to collect a new set of trajectories for the next round of training. However, when having large gradient updates in training, the process becomes unstable. To address this, researchers proposed Trust Region Policy Optimization (TRPO) \cite{schulman2015trust} and Proximal Policy Optimization (PPO) \cite{schulman2017proximal} to further improve the stability of training.

\subsubsection{Actor-Critic Network}

\begin{figure}[ht]
    \centering
    \includegraphics[width=0.9\linewidth]{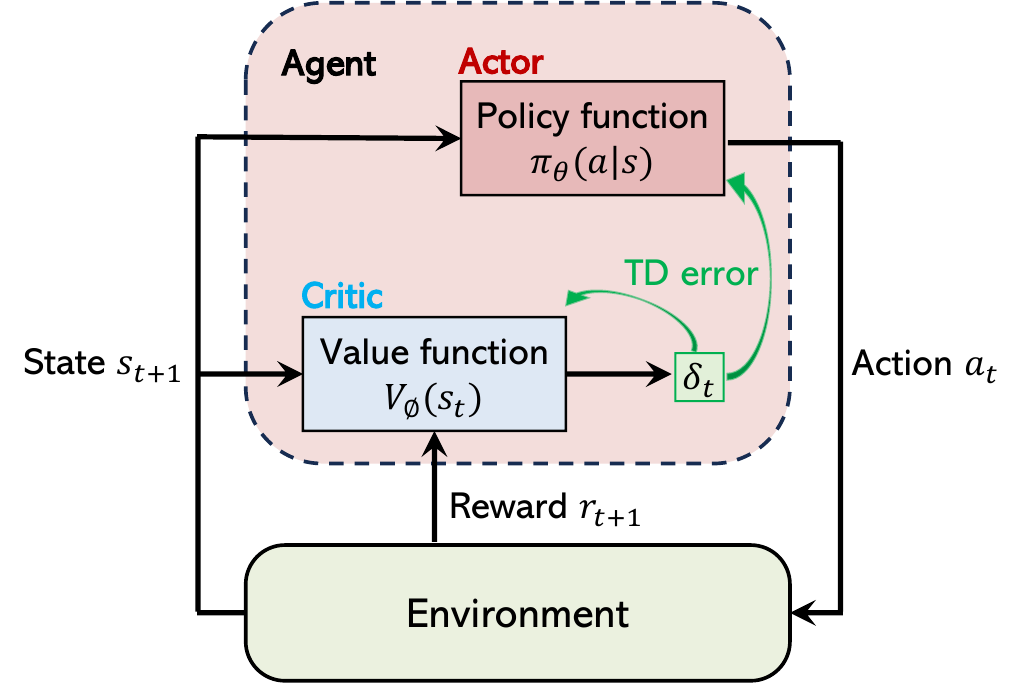} 
    \caption{Classic structure of an Actor-Critic network.}
    \label{ac}
\end{figure}

Actor-Critic \cite{konda1999actor} is one of the major DRL algorithms. The \textit{Actor} learns the policy to take action within a given state directly (the hyper-parameters set is denoted as $\theta$), which is different from DQN (indirectly learning by approaching optimal Q value). The \textit{critic} is utilized to evaluate the value of the policy by approaching an accurate value function using a neural network (Hyper-parameters set is denoted as $\phi$).

Actor and critic are initialized by the $\theta_0$ and $\phi_0$. Then, based on the initial policy $\pi_{\theta_0}(a|s)$, actor starts to explore the environment to collect interaction experience $(s_t, a_t, r_t, s_{t+1})$. Critic, with initial $\phi_0$ parameters, aims to evaluate the value of the  policy by calculating temporal difference (TD) in \Cref{TD_difference} using actor’s experience. The smaller $\mathbb{E}(\delta_t)$ is, the better $\pi_{\theta}$ is. Furthermore, $\phi_0$ will be updated using $\delta_t$ and gradient descent in \Cref{gd_10}.

\begin{equation}
\delta_t = r_t + \gamma V_{\phi}(s_{t+1}) - V_{\phi}(s_t)
\label{TD_difference}
\end{equation}

\begin{equation}
\phi \leftarrow \phi + \alpha_\phi \delta_t \nabla_{\phi} V_{\phi}(s_t)
\label{gd_10}
\end{equation}

After updating the Critic, the Actor parameters are updated based on the TD error $\delta_t$, which serves as an estimate of the advantage function. The policy gradient is computed as shown in \Cref{actor_gradient}, and the Actor parameters are updated via gradient ascent: $\theta \leftarrow \theta + \alpha_\theta \nabla_{\theta} J(\theta)$, where $\alpha_\theta$ is the learning rate for the Actor.

\begin{equation}
\nabla_{\theta} J(\theta) = \mathbb{E}_{(s_t,a_t) \sim \pi_\theta}[\nabla_{\theta} \log \pi_{\theta}(a_t|s_t) \cdot \delta_t]
\label{actor_gradient}
\end{equation}

It is worth noting that the TD error $\delta_t$ serves as an unbiased estimate of the advantage function, representing the relative benefit of action $a_t$ compared to the baseline value $V(s_t)$.

Based on the idea of the actor-critic network, many improved versions have been proposed. Advantage Actor-Critic (A2C) \cite{mnih2016asynchronousmethodsdeepreinforcement} is an improved version of the basic Actor-Critic method that incorporates the \textit{advantage function} $A(s, a)$ to reduce variance in gradient estimates. The advantage function evaluates the relative benefit of choosing a specific action $a$ in state $s$ compared to the expected value under the current policy, thereby improving training stability and efficiency. Based on this idea, the update for \textit{actor} in A2C is followed by \Cref{A2Cupdate} .

\begin{equation}
A(s, a) = Q(s, a) - V(s)
\end{equation}

\begin{equation}
Q(s_t, a_t) \approx r_t + \gamma V(s_{t+1}; \phi)
\label{Q_approx}
\end{equation}

\begin{equation}
\nabla_\theta J(\theta) = \mathbb{E}_{(s_t,a_t) \sim \pi_\theta}\left[\nabla_\theta \log \pi_\theta(a_t | s_t) \cdot A(s_t, a_t)\right]
\label{A2Cupdate}
\end{equation}

\begin{equation}
L(\phi) = \mathbb{E}_{(s_t,r_t,s_{t+1}) \sim \pi_\theta}\left[(r_t + \gamma V(s_{t+1}; \phi) - V(s_t; \phi))^2\right]
\end{equation}

The aim of Critic in A2C is making sure to predict the expected return given a specific state $s$ following current policy  $\pi$ . Thus, trying to minimize the difference between prediction of value function and target return is essential to update the parameters of critic, shown in the \Cref{ac}. Asynchronous Advantage Actor-Critic (A3C) \cite{mnih2016asynchronousmethodsdeepreinforcement} parallelize the learning process across multiple asynchronous threads (workers). Each worker maintains its own copy of the environment and independently collects experience using the shared global network parameters. After computing local gradients over multiple steps, each worker asynchronously updates the shared global parameters using standard gradient descent, without requiring synchronization among workers. This asynchronous approach improves training efficiency and stability through diverse exploration \cite{sewak2019actor}.

\subsection{Inverse reinforcement learning}
Inverse Reinforcement Learning (IRL) constitutes a problem setting within RL, aiming to infer the reward function from observed expert behaviours \cite{arora2021survey}. Unlike traditional reinforcement learning, where agents learn optimal policy through interactions with the environment based on a predefined reward function, IRL focuses on understanding and replicating such behaviours without directly knowing the reward function, by observing exemplary policies or actions \cite{ng2000algorithms}. The fundamental premise of IRL is that the observed behaviours reflect the intrinsic motivations or reward structures adhered to during these actions. Consequently, by inversely inferring these motivations or rewards, IRL seeks to construct a reward function that can explain the observed behaviours and can be used to guide agents in learning similar strategies. The most widely used IRL methods include Maximum Entropy Inverse Reinforcement Learning, which uses a linear function to approximate the reward function behind \cite{ziebart2008maximum}. However, for the complex reward function, the method holds limitations. Based on that, Maximum  Entropy Deep Inverse Reinforcement Learning was proposed, using fully convolutional neural networks to represent the reward function \cite{wulfmeier2015maximum}.

\subsection{Evaluation}
Evaluations in RL are primarily concentrated on assessing the performance of agent in particular task or environment \cite{jordan2020evaluating}. It relies on specific purpose of tasks, goals of agent and any available feedback information. There are 2 mainly used evaluation metrics, cumulative reward or discounted cumulative reward (return), which means total reward that the agent could gain in one episode or within a fixed period. Success rate measures how often the agent reaches its specific goals or successfully completes assigned tasks. It is applicable for tasks with clear success criteria, e.g. navigation. Other evaluation metrics could be set up by understanding the final purpose of task as well, it will depend on the scenario where RL applied. Specifically, in network intrusion detection scenarios, accuracy, \hyperref[precision]{precision}, \hyperref[recall]{recall} and \hyperref[F1]{$F_1$ scores} are widely utilized, by calculating True Positive (TP), False Negative (FN), False Positive (FP) and True Negative (TN) from confusion matrix. In some cases, Receiver Operating Characteristics (ROC) curve \cite{hanley1989receiver} will be used to measure the model performances. The Area Under the ROC Curve (AUC) shown in \Cref{auc} quantifies overall discriminative ability by integrating the true positive rate over the false positive rate as the decision threshold $t$ varies, where $TPR(t) = \frac{TP(t)}{TP(t)+FN(t)}$ and $FPR(t) = \frac{FP(t)}{FP(t)+TN(t)}$ denote the true positive rate and false positive rate as functions of the decision threshold $t$, respectively.

\begin{table*}[t]
\caption{List of tools and their accessible links for deep reinforcement learning based network intrusion.}
\begin{tabular*}{\textwidth}{lll}
\hline
\textbf{Name of Tools} & \textbf{Brief Introduction} & \textbf{Accessible here} \\ \hline
CSLE & A platform for evaluating and developing RL agents for control problems. & \href{https://github.com/Limmen/csle}{Limmen/csle}  \\
PenGym & A Penetration testing framework for creating and managing real-world environments. & \href{https://github.com/cyb3rlab/PenGym}{cyb3rlab/PenGym} \\
AutoPen & An automated penetration testing framework. & \href{https://github.com/crond-jaist/AutoPentest-DRL}{crond-jaist/AutoPentest-DRL} \\
NASimEmu &  A framework for training agents in offensive penetration-testing scenarios. & \href{https://github.com/jaromiru/NASimEmu}{jaromiru/NASimEmu} \\
CLAP & A simulated computer network complete with vulnerabilities, scans and exploits. & \href{https://github.com/yyzpiero/RL4RedTeam}{yyzpiero/RL4RedTeam} \\
Cyberwheel & A simulation environment focused on autonomous cyber defence. & \href{https://github.com/ORNL/cyberwheel}{ORNL/cyberwheel} \\
Idsgame & A environment for simulating attack and defence operations. & \href{https://github.com/Limmen/gym-idsgame}{Limmen/gym-idsgame} \\
MAB-Malware & An open-source framework to generate specific malware.  & \href{https://github.com/weisong-ucr/MAB-malware}{weisong-ucr/MAB-malware} \\
YAWNING-TITAN & An abstract, graph based cyber-security simulation environment. & \href{https://github.com/dstl/YAWNING-TITAN}{dstl/YAWNING-TITAN} \\
PrimAITE & PrimAITE is a configurable reinforcement learning simulation environment. & \href{https://github.com/Autonomous-Resilient-Cyber-Defence/PrimAITE}{PrimAITE} \\ \hline
\end{tabular*}
\label{useful_RL_tools}
\end{table*}

\begin{equation}
Accuracy = \frac{TP + TN}{TP + TN + FP + FN}
\end{equation}

\begin{equation}
Precision = \frac{TP}{TP + FP}
\label{precision}
\end{equation}

\begin{equation}
Recall = \frac{TP}{TP + FN}
\label{recall}
\end{equation}

\begin{equation}
F_1 = 2 \times \frac{Precision \times Recall}{Precision + Recall}
\label{F1}
\end{equation}

\begin{equation}
\label{auc}
    AUC = \int_{0}^{1} TPR(t) \, d\,FPR(t)
\end{equation}

\section{Tools used for RL based network intrusion}
For accelerating the evaluation process of DRL models in network intrusion issues, researchers have developed many useful tools including CSLE \cite{hammar2022intrusion}, PenGym \cite{nguyen2024pengym}, AutoPen \cite{hu2021automated}, NASimEmu \cite{janisch2023nasimemu}, CLAP \cite{yang2022behaviour}, Cyberwheel \cite{oesch2024towards}, Idsgame \cite{hammar2020finding}, MAB-Malware \cite{song2020mab} and YAWNING-TITAN \cite{andrew2022developing}, they are shown in \Cref{useful_RL_tools}. Here are some basic information to introduce what and how will these tools impact the DRL-based network intrusion detection.

CSLE is a platform designed for developing and testing reinforcement learning agents in network intrusion detection, offering a realistic cyber range environment. It supports integration with methods like dynamic programming, game theory, and optimization, enhancing cyber security research. PenGym is a framework for training RL agents in penetration testing, compatible with the Gymnasium API. It allows RL agents to perform actions like network scanning and exploitation in controlled environments. AutoPen is an automated penetration testing framework that uses DRL to identify optimal attack paths in both simulated and real networks. It integrates tools like Nmap \cite{lyon2009nmap} and Metasploit\cite{raj2020study} to execute attacks, allowing users to study penetration testing techniques for educational purposes. NASimEmu is a framework designed for training DRL agents in offensive penetration-testing scenarios, featuring both a simulator and an emulator for seamless deployment. It uses a random generator to create varied network scenarios and supports simultaneous training across multiple scenarios. CLAP is a RL agent based on PPO that performs penetration testing in simulated network environments using the Network Attack Simulator (NASim). The agent is trained to identify and exploit vulnerabilities to gain access to network resources. Cyberwheel is a RL simulation environment designed for training and evaluating autonomous cyber defence models on simulated networks. Built with modularity, it allows users to customize networks, services, host types, and defensive agents through configurable files. Idsgame is a RL environment designed for simulating attack and defence operations within an abstract network intrusion game. Based on a two-player Markov game model, it features attacker and defender agents competing in a simulated network. The environment provides an interface to a partially observed Markov decision process (POMDP), enabling the training, simulation, and evaluation of attack and defence policies. MAB-Malware is an open-source reinforcement learning framework designed to generate adversarial examples for PE malware by modeling the problem as a multi-armed bandit (MAB). Each action-content pair is treated as an independent slot machine with rewards modeled by a Beta distribution, and Thompson sampling is used to balance exploration and exploitation. YAWNING-TITAN (YT) is a graph-based cyber-security simulation environment built to train intelligent agents for autonomous cyber defence operations. It focuses on simplicity, minimal hardware requirements, and is platform-independent, supporting various algorithms and customizable environment settings.

In the end, PrimAITE (Primary-level AI Training Environment) is a simulation platform designed for training and evaluating AI agents in cyber defense scenarios. The environment features high configurability through YAML files, realistic network traffic simulation, and support for multiple agents with customizable observation and action spaces. It operates at machine speed to enable rapid training cycles and incorporates reinforcement learning reward functions based on threat mitigation and mission success. PrimAITE ships with example scenarios but is designed as a flexible framework that users can extend and reconfigure to address their specific cyber defense training needs.

\section{DRL in Network Intrusion detection}

DRL applied to NID has shown remarkable capabilities in the past years \cite{nguyen2021deep, alavizadeh2022deep, rizzardi2023deep, vadigi2023federated, faker2019intrusion} as demonstrated by the rapid increase presented in \Cref{trend} . By combining deep learning and RL, researchers can develop highly efficient, automatic learning detection models. These studies highlight DRL’s powerful ability to process high-dimensional data and complex network environments \cite{faker2019intrusion}. It can not only improve detection accuracy and sensitivity, but also optimize its performance through a continuous and dynamic learning process, making network intrusion defence more intelligent and automated \cite{nguyen2021deep}. These advantages suggest that DRL will play an increasingly important role in future NID research. In the following subsections, we aim to fully investigate current research and applications for last 5 years and present insightful views on deep reinforcement learning for network intrusion detection.

\subsection{Dataset}
\begin{figure*}[ht]
    \centering
    \includegraphics[width=\textwidth]{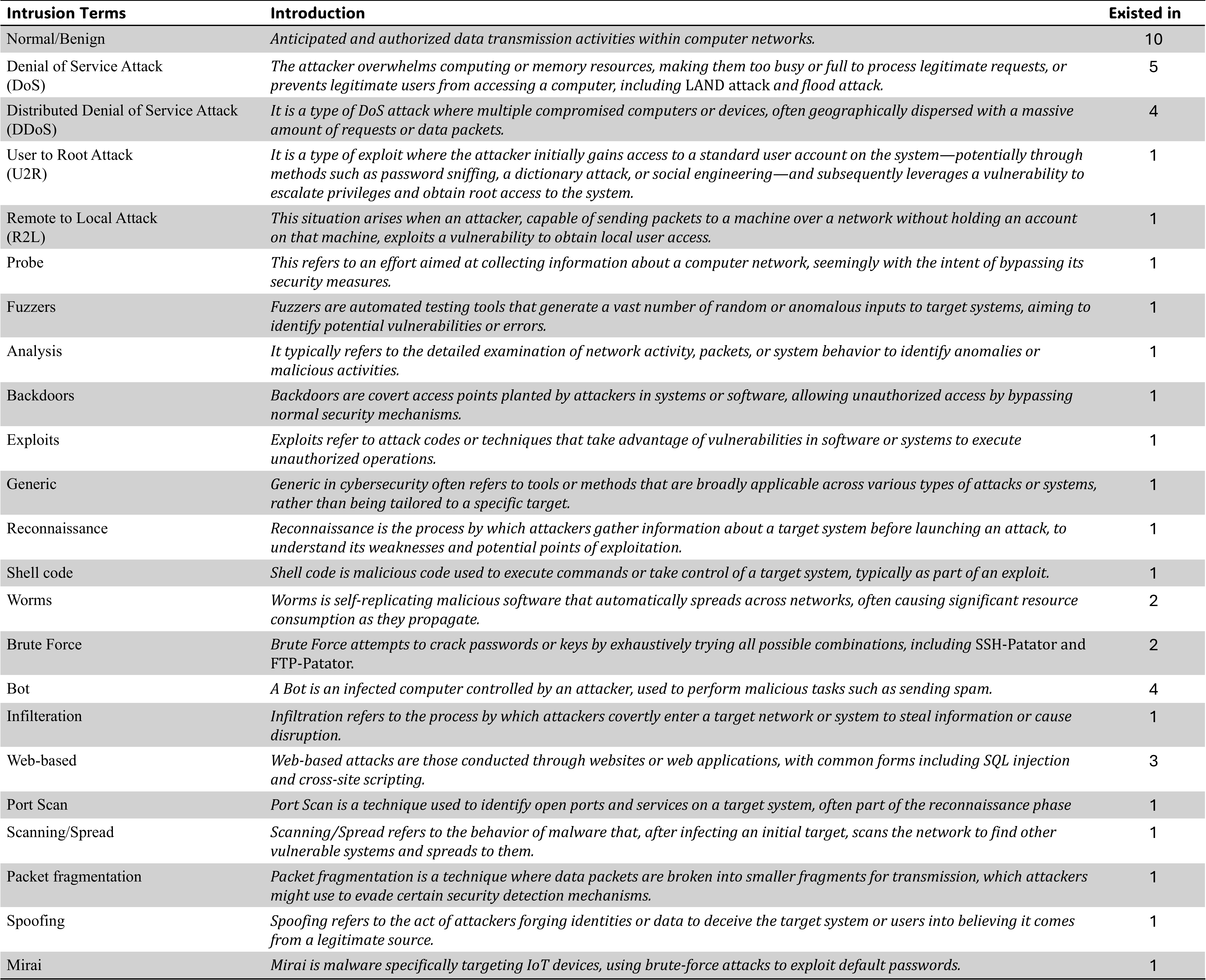}
    \caption{An introduction to frequently mentioned intrusion terms and their frequency across the datasets considered.}
    \label{attckdefinitions}
\end{figure*}

Many intrusion datasets have been proposed in recent years. Attack or intrusion types are various in different intrusion datasets. In this section, we are going to present the basic information of current widely utilised network intrusion datasets including \href{https://kdd.ics.uci.edu/databases/kddcup99/kddcup99.html}{KDDCUP 99}, \href{https://ieee-dataport.org/documents/nsl-kdd-0}{NSL-KDD} \href{https://kilthub.cmu.edu/articles/dataset/Insider_Threat_Test_Dataset/12841247}{CMU-CERT}, \href{https://research.unsw.edu.au/projects/unsw-nb15-dataset}{UNSW-NB15}, \href{https://www.unb.ca/cic/datasets/ids-2017.html}{CIC-IDS-2017}, \href{https://www.unb.ca/cic/datasets/ids-2018.html}{CSE-CIC-IDS2018}, \href{https://www.unb.ca/cic/datasets/ddos-2019.html}{CICDDoS2019}, \href{https://github.com/Grigaliunas/electronics9050800/}{LITNET-2020}, \href{https://icsdweb.aegean.gr/awid/download-dataset}{AWID}, \href{https://secplab.ppgia.pucpr.br/?q=reinforcemawiflow}{MAWIFlow}, \href{https://www.unb.ca/cic/datasets/iotdataset-2023.html}{CICIoT2023}, providing essential understanding including distributions of intrusion types and network traffic features in each datasets and how to collect or create them. In the end, a comprehensive summary for all datasets are presented for any future reference. The introduction of all mentioned intrusions in the following datasets is presented in \Cref{attckdefinitions} for reference.

The KDDCUP 99 dataset \cite{tavallaee2009detailed} was created by processing data from the 1998 DARPA \cite{mchugh2000testing} intrusion detection challenge. This was achieved using the Mining Audit Data for Automated Models for Intrusion Detection (MADMAID) \cite{lee1998mining,nazer2012evaluating} framework to extract features from the raw tcpdump data. Detailed statistics of the dataset are provided in \Cref{KDDCup_NSL}. The original 1998 dataset was developed by MIT's Lincoln Laboratory, involving thousands of UNIX machines and hundreds of users. Network traffic was recorded in tcpdump format \footnote{The tcpdump format is used to capture detailed network traffic information, including timestamps, source and destination IP addresses, port numbers, protocol types, and more.} over a ten-week period, with the first seven weeks' data serving as the training set and the remaining three weeks' data as the testing set. The KDDCUP 99 DARPA dataset is available in two versions: the full dataset and a 10\% sample. It includes 41 features and is categorized into five classes: \textit{Normal, DoS, Probe, R2L, and U2R}. 

The NSL-KDD dataset \cite{425a-3e55-18} is one of the most widely recognized benchmark datasets, extensively utilized by cyber-security researchers to evaluate the performance of Intrusion Detection Systems (IDS). Developed by \cite{tavallaee2009detailed}, it is based on the KDDCUP 99 dataset encompassing both attack and non-attack instances. It is classified as either normal or one of 38 predefined attack types. The training subset includes 22 specific attack types, while the testing subset introduces an additional 16 novel attack types. It retains the original four types of attacks from the KDDCUP 99 dataset.

\begin{table}[ht]
    \centering
    \caption{Intrusions distribution for KDDCup 99 and NSL-KDD datasets}
    \begin{tabular}{lrrrr}
        \toprule
        \multirow{2}{*}{\textbf{Intrusions}} & \multicolumn{2}{c}{\textbf{KDDCup 99}} & \multicolumn{2}{c}{\textbf{NSL-KDD}} \\
        \cline{2-5}
        & \textbf{Train Set} & \textbf{Test Set} & \textbf{Train Set} & \textbf{Test Set} \\
        \midrule
        Normal & 97,278 & 60,593 & 67,343 & 9,710 \\
        DoS & 391,458 & 229,853 & 45,927 & 7,458 \\
        Probe & 4,107 & 4,166 & 11,656 & 2,422 \\
        R2L & 1,126 & 16,189 & 995 & 2,887 \\
        U2R & 52 & 228 & 52 & 67 \\
        \midrule
        Total records & 494,021 & 311,029 & 125,973 & 22,544 \\
        \bottomrule
    \end{tabular}
    \label{KDDCup_NSL}
\end{table}

CMU-CERT \cite{trzeciak2011cert} is a synthetic dataset firstly proposed by Computer Emergency and Response Team (CERT) division of Carnegie Mellon University (CMU) as an insider threat dataset. Additionally, this dataset has been continuously updated in recent years. However, one significant drawback is the substantial data imbalance \cite{singh2021user}. The insider threat dataset is available in several iterations in recent years, with each new version offering enhancements and improvements. Version 4.2 is notably more frequently utilized, as it includes the highest proportion of intrusions relative to normal data. This version comprises 30,602,325 entries in total, of which 7,623 entries are identified as attacks. Consequently, the percentage of intrusions in this version is approximately 0.025\%. The dataset encompasses two years' worth of Lightweight Directory Access Protocol (LDAP) \footnote{Accessing and managing distributed directory information services.} logs, which are instrumental in pinpointing the active users within the company at any given moment. The insider threat dataset, involving 70 employees, is based on three specific scenarios. 

\begin{itemize}
  \item Scenario 1: An employee with no prior record of using removable drives or working after hours suddenly begins logging in outside of business hours, utilizing a removable drive, and uploading data to \textit{Wikileaks.org}. This activity is followed by the user's swift departure from the company. 
  \item Scenario 2: An employee begins to visit job search websites and applies for positions at competing firms. Prior to leaving the organization, this individual uses a thumb drive to steal a significant amount of data, far exceeding their previous usage. 
  \item Scenario 3: A disgruntled system administrator installs a keylogger and transfers it to his supervisor's computer via a thumb drive. The next day, he exploits the captured keystrokes to log in as his supervisor and sends a panic-inducing mass email to the entire organization, before immediately resigning.
\end{itemize}

\begin{table}[ht]
    \centering
    \caption{Data Sources in the CMU-CERT Insider Threat Dataset V4.2}
    \begin{tabular}{lrrr}
        \toprule
        \textbf{Sources} & \textbf{Total Entries} & \textbf{Intrusions} & \textbf{Percentage(\%)} \\
        \midrule
        Logon & 427,628 & 198 & 0.046 \\
        Device & 205,476 & 2,786 & 1.37 \\
        HTTP & 28,438,284 & 3,860 & 0.013 \\
        Email & 1,315,459 & 469 & 0.035 \\
        File & 222,801 & 10 & 0.004 \\
        \bottomrule
    \end{tabular}
    \label{CMU-CERT}
\end{table}

To address the issues including redundant records, imbalanced datasets and too many simple records present in the KDDCup 99 and NSL-KDD datasets, the research team at the Australian Centre for Cyber Security (ACCS) developed a new dataset known as UNSW-NB15 \cite{moustafa2015unsw}. This dataset was created using a hybrid generation method, employing the IXIA Perfect Storm tool \footnote{A network security testing tool primarily utilized for evaluating and testing the security performance of network infrastructure.} to capture real-time network traffic containing both normal and malicious activities. The IXIA Perfect Storm tool includes a library that stores new attacks and common vulnerabilities and exposures (CVEs) \footnote{Common vulnerabilities and exposures at \url{https://www.cve.org/}}, which is a publicly available repository of security vulnerabilities and exposures.

During the data generation process, the researchers utilized two servers: one to simulate normal network activities and the other to generate malicious activities. The network data packets were captured using the tcpdump tool. The entire process took 31 hours, resulting in the collection of 100 GB of data, which was subsequently divided into multiple 1,000 MB pcap files. Following this, the researchers used Argus and Bro-IDS on a Linux Ubuntu 14.0.4 system to extract features from these pcap files. Additionally, they developed 12 algorithms to conduct an in-depth analysis of each data packet \cite{moustafa2015unsw}. Finally, the UNSW-NB15 dataset is presented as full connection records: comprising 2 million connection records and partial connection records: consisting of 82,332 training records and 175,341 testing records, covering 10 types of attacks. Notably, the partial connection record dataset contains 42 features and includes corresponding class labels that categorize network behaviours into normal and nine different types of attacks.

\begin{table}[ht]
    \centering
    \caption{Intrusion types and their respective train and test data distributions for UNSW-NB15 dataset.}
    \begin{tabular}{lrrrr}
        \toprule
        \textbf{Intrusions} & \textbf{Train} & \textbf{Ratio(\%)} & \textbf{Test} & \textbf{Percentage(\%)} \\
        \midrule
        Normal         & 56,000  & 60.22 & 37,000  & 39.78 \\
        Fuzzers        & 18,184  & 75.00 & 6,062   & 25.00 \\
        Analysis       & 2,000   & 74.71 & 677    & 25.29 \\
        Backdoors      & 1,746   & 74.97 & 583    & 25.03 \\
        DoS            & 12,264  & 75.00 & 4,089   & 25.00 \\
        Exploits       & 33,393  & 75.00 & 11,132  & 25.00 \\
        Generic        & 40,000  & 67.95 & 18,871  & 32.05 \\
        Reconnaissance & 10,491  & 75.01 & 3,496   & 24.99 \\
        Shell code     & 1,133   & 74.98 & 378    & 25.02 \\
        Worms          & 130    & 74.71 & 44     & 25.29 \\
        \midrule
        Total          & 17,5341 & 68.05 & 82,332  & 31.95 \\
        \bottomrule
    \end{tabular}
    \label{UNSW-NB15}
    
\end{table}

In CIC-IDS-2017 \cite{ids2017datasets}, researchers utilized CICFlowMeter tool to extract 80 network traffic features based on a 5-day traffic flow data. The principal objective during data collection was to capture authentic background traffic. Employing the B-profile system, benign traffic was characterized by 25 users operating across protocols including HTTP, HTTPS, FTP, SSH, and email. The data collection extended over a period of five days, documenting routine traffic on one day and introducing various attacks on subsequent days. The injected attacks encompassed Brute Force FTP, Brute Force SSH, DoS, Heartbleed, Web Attack, Infiltration, Botnet, and DDoS.

\begin{table}[ht]
    \centering
    \caption{Attack distributions in train and test sets for CIC-IDS-2017}
    \begin{tabular}{lrrrr}
        \toprule
        \textbf{Intrusion Types} & \textbf{Train} & \textbf{Percentage(\%)} & \textbf{Test} & \textbf{Percentage (\%)} \\
        \midrule
        Normal        & 60,000  & 75.00 & 20,000 & 25.00 \\
        SSH-Patator   & 5,000   & 84.79 & 897    & 15.21 \\
        FTP-Patator   & 7,000   & 88.18 & 938    & 11.82 \\
        DoS           & 6,000   & 75.00 & 2,000  & 25.00 \\
        Web           & 2,000   & 91.74 & 180    & 8.26 \\
        Bot           & 1,500   & 76.30 & 466    & 23.70 \\
        DDoS          & 6,000   & 75.00 & 2,000  & 25.00 \\
        PortScan      & 6,000   & 75.00 & 2,000  & 25.00 \\
        \midrule
        Total          & 93,500  & 76.65 & 28,481 & 23.35 \\
        \bottomrule
    \end{tabular}
    \label{CIC-IDS-2017}
\end{table}

The CSE-CIC-IDS2018 dataset \cite{ids2018datasets} is among the most widely used for IDS. It encompasses seven distinct attack scenarios: Brute-force, Heartbleed, Botnet, DoS, DDoS, Web assaults, and internal network penetration. The attacking infrastructure comprises 50 machines, while the victim organization’s infrastructure includes 420 machines and 30 servers spread across five different departments. This dataset features network traffic and system logs for each computer, with up to 84 network features extracted from the recorded network traffic using CICFlowMeter-V3 as well. What is worth mentioning is that the dataset exhibits an unequal distribution of positive and negative samples, typically addressed through over-sampling or down-sampling to manage the imbalance.

\begin{table}[ht]
    \centering
    \caption{CSE-CIC-IDS2018 dataset intrusion distribution.}
    \begin{tabular}{lrr}
        \toprule
        \textbf{Intrusion Types} & \textbf{Traffic Flow Counting} & \textbf{Percentage (\%)} \\
        \midrule
        Benign        & 13,484,708 & 83.07 \\
        DDoS          & 1,263,933  & 7.79 \\
        DoS           & 654,300    & 4.03 \\
        Brute Force   & 380,949    & 2.35 \\
        Bot           & 286,191    & 1.76 \\
        Infiltration & 161,934    & 0.99 \\
        Web           & 928        & 0.01 \\
        \bottomrule
    \end{tabular}
    \label{CSE-CIC-IDS2018}
\end{table}

The CICDDoS2019 dataset \cite{sharafaldin2019developing}, developed by the Canadian Institute for Cybersecurity (CIC) at the University of New Brunswick (UNB), serves as a realistic and comprehensive benchmark for the detection of Distributed Denial of Service (DDoS) attacks. This dataset addresses the shortcomings of existing datasets, such as incomplete traffic, anonymous data, and outdated attack scenarios. It encompasses 11 distinct types of DDoS attacks, including reflective and exploitative attacks, with 80 network traffic features extracted and calculated from all benign and denial-of-service flows using the CICFlowMeter software. Additionally, the dataset is generated by simulating a real-world network environment, incorporating genuine interactions between the attacker and victim networks, as well as attacks executed using third-party tools and packages. The attack distributions of CICDDoS2019 are presented in \cref{CICDDoS2019}.

\begin{table}[ht]
    \centering
    \caption{CICDDoS2019 dataset intrusion distributions.}
    \begin{tabular}{lrr}
        \toprule
        \textbf{Intrusion Types} & \textbf{Traffic Flow Counting} & \textbf{Percentage (\%)} \\
        \midrule
        Benign        & 56,863      & 0.11 \\
        DDoS DNS      & 5,071,011   & 10.13 \\
        DDoS LDAP     & 2,179,930   & 4.35 \\
        DDoS MSSQL    & 4,522,492   & 9.03 \\
        DDoS NetBIOS  & 4,093,279   & 8.18 \\
        DDoS NTP      & 1,202,642   & 2.40 \\
        DDoS SNMP     & 5,159,870   & 10.31 \\
        DDoS SSDP     & 2,610,611   & 5.21 \\
        DDoS SYN      & 1,582,289   & 3.16 \\
        DDoS TFTP     & 20,082,580  & 40.11 \\
        DDoS UDP      & 3,134,645   & 6.26 \\
        DDoS UDP-Lag  & 366,461     & 0.73 \\
        \midrule
        Total         & 50,062,673  & 100.00 \\
        \bottomrule
    \end{tabular}
    \label{CICDDoS2019}
\end{table}

LITNET-2020 \cite{damasevicius2020litnet} is a novel benchmark dataset for network traffic based intrusion detection proposed by Kaunas University of Technology, Lithuania. This dataset is designed to provide realistic and up-to-date network traffic data for the development of Network Intrusion Detection (NID) methods. Despite numerous recent efforts that have introduced various benchmark datasets for NID, existing datasets still fail to adequately capture modern network traffic scenarios and provide examples of diverse network attacks and intrusions. LITNET-2020 fills this gap by offering annotated data obtained from a real-world academic network. Captured from the Lithuanian Research and Education Network (LITNET) between March 6, 2019, and January 31, 2020, the dataset comprises over 45,330,333 records, with 5,328,934 records being attack data, covering 12 types of attacks. The dataset includes 49 attributes from the NetFlow v9 protocol and is extended with 19 custom attack detection features, with each record containing 85 network traffic features. LITNET-2020 provides a valuable resource for researchers in the field of cybersecurity, aiding in the development and validation of more effective NIDs \Cref{LITNET-2020} shows the intrusion distribution of LITNET-2020

\begin{table}[ht]
    \centering
    \caption{Intrusion distribution of LITNET-2020 dataset.}
    \begin{tabular}{lrrr}
        \toprule
        \textbf{Intrusion Types} & \textbf{Total} & \textbf{Attacks} & \textbf{Percentage (\%)} \\
        \midrule
        Smurf                   & 3,994,426  & 59,479    & 1.49 \\
        ICMP-flood              & 3,863,655  & 11,628    & 0.30 \\
        UDP-flood               & 606,814    & 93,583    & 15.42 \\
        TCP SYN-flood           & 14,608,678 & 3,725,838 & 25.50 \\
        HTTP-flood              & 3,963,168  & 22,959    & 0.58 \\
        LAND attack             & 3,569,838  & 52,417    & 1.47 \\
        Blaster Worm            & 2,858,573  & 24,291    & 0.85 \\
        Code Red Worm           & 5,082,952  & 1,255,702 & 24.70 \\
        Spam bot’s detection    & 1,153,020  & 747       & 0.06 \\
        Reaper Worm             & 4,377,656  & 1,176     & 0.03 \\
        Scanning/Spread         & 6,687      & 6,232     & 93.20 \\
        Packet fragmentation    & 1,244,866 & 477     & 0.04 \\
        \midrule
        Total flows                   & 45,330,333 & 5,328,934 & 11.76 \\
        \bottomrule
    \end{tabular}
    \label{LITNET-2020}
\end{table}

The Aegean WiFi Intrusion Dataset (AWID) \cite{kolias2015intrusion, chatzoglou2022pick} is a publicly accessible and comprehensive dataset meticulously crafted for research in wireless network security and intrusion detection. It encapsulates traffic captured from actual 802.11 networks secured via WEP encryption, encompassing both benign and adversarial traffic. The dataset is bifurcated into two principal variants: the AWID-ATK, delineated by attack types, and the AWID-CLS, categorized by attack classes. Each variant is enhanced with both full and reduced subsets, accompanied by their respective training and testing sets. Within the dataset, each packet is represented as a vector comprising 156 attributes, including but not limited to the Source Address, Destination Address, Initialization Vector (IV), Extended Service Set Identifier (ESSID), and Signal Strength. These attributes have been subjected to preprocessing, and transformed into numerical or categorical values, thereby facilitating the analysis via machine learning algorithms. It is already a benchmark used for the research and development of wireless IDS, focusing on the security of 802.11, which is a specific wireless frequency.

The MAWIFlow dataset \cite{viegas2019bigflow} is a publicly available compilation of around 8 terabytes of real network traffic spanning a four-year interval from 2016 to 2019, encompassing more than seven billion network flows. It ensures authenticity and diversity by including a multitude of protocols and behaviours. Labelling of daily anomalous events is facilitated through the MAWILab tool, which identifies various network-level attack types such as Service Scan, TCP Scan, and Denial-of-Service attacks. The dataset's high variability and completeness are further augmented by the application of the BigFlow feature extraction algorithm, which extracts a set of thirty-nine features for each network flow within a 15-second time window, making it an ideal benchmark for evaluating the performance and model update strategies of intrusion detection techniques in the face of evolving network traffic behaviours.

\begin{table}[ht]
    \centering
    \caption{Intrusion distributions of CICIoT2023 datasets.}
    \begin{tabular}{lrr}
        \toprule
        \textbf{Intrusion Types} & \textbf{Traffic Flow Counting} & \textbf{Percentage (\%)} \\
        \midrule
        DDoS        & 33,984,650 & 74.55 \\
        DoS         & 8,090,738  & 17.75 \\
        Recon       & 354,565    & 0.78 \\
        Web-Based   & 24,829     & 0.05 \\
        Spoofing    & 499,568    & 1.10 \\
        Mirai       & 2,634,124  & 5.78 \\
        \midrule
        Total       & 45,588,474 & 100.00 \\
        \bottomrule
    \end{tabular}
    \label{CIC2023}
\end{table}

The CICIoT2023 dataset \cite{neto2023ciciot2023}, shown in \cref{CIC2023}, is a novel and extensive Internet of Things (IoT) attack dataset, designed to foster the development of security analytics applications in real IoT operations. This dataset is generated by executing 33 types of attacks within an IoT topology comprising 105 devices, categorized into seven distinct classes, namely Distributed Denial of Service (DDoS), Denial of Service (DoS), Reconnaissance (Recon), Web-based attacks, Brute Force, Spoofing, and Mirai. All attacks are conducted by malicious IoT devices targeting other IoT devices. The dataset encompasses a variety of IoT device types, such as smart home devices, cameras, sensors, and micro-controllers. Forty-seven network traffic features are extracted from the dataset, including packet length, transmission rate, and protocol types. It is available in two file formats, pcap and csv, to accommodate the needs of researchers. The pcap files contain the raw data, while the csv files provide the extracted features.

\subsection{Modeling workflow}
\begin{figure}[ht]
    \centering
    \includegraphics[width=0.75\linewidth]{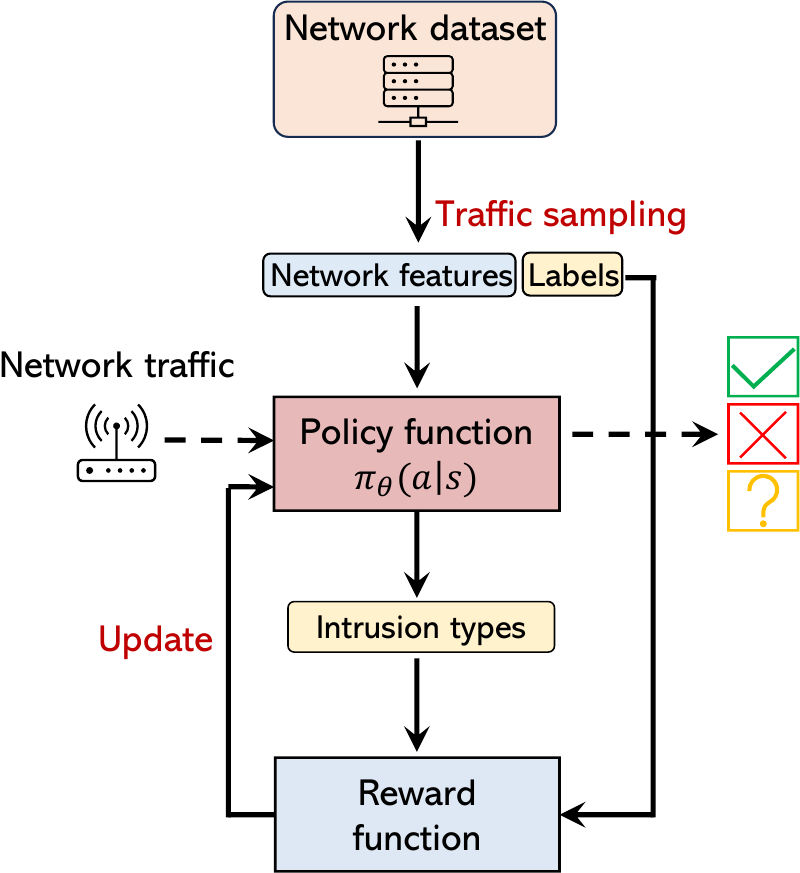} 
    \caption{Deep reinforcement learning in network intrusion detection.}
    \label{NID_DRL}
\end{figure}

\Cref{NID_DRL} illustrates the current standard process of implementing a network intrusion detection model using DRL. In the first phase, a network dataset is obtained, then, by employing a traffic sampling method, it is possible to extract network features and labels which are needed/utilized in the training phase. In this phase, the policy function \( \pi_{\theta}(a|s) \) receives network traffic features data as input and predicts potential intrusion types. Based on these predictions, feedback is provided in the form of correct, incorrect, or uncertain detection outcomes. This feedback is further processed through a reward function, which in turn updates the policy function to enhance its detection accuracy. Designing a practical reward function could accelerate the training speed and the aim of training is to let the policy function get the maximum reward. During the testing phase, the trained policy function is applied to real network traffic, producing intrusion detection results. This entire process iteratively updates the policy function, enabling the system to more effectively identify and classify network intrusions in a dynamic network environment.

\subsection{Model performance}

\begin{table*}[ht]
    \centering
    \caption{Deep reinforcement learning model performances on NSL-KDD datasets in last few years.}
    \begin{tabular*}{\textwidth}{@{\extracolsep{\fill}} llllcclllll}
        \toprule
        \textbf{Reference} & \textbf{Year} & \textbf{Method} & \textbf{Dataset} & \multicolumn{4}{c}{\textbf{Best Model Performance}} \\
        \cmidrule(lr){5-8}
        & & & & \textbf{ACC(\%)} & \textbf{PR(\%)} & \textbf{RC(\%)} & \textbf{F1(\%)} \\
        \midrule
        \cite{louati2024big} & 2024 & Big-IDS & NSL-KDD & 97.44 & / & / & / \\
        \cite{li2023soft} & 2023 & AE-SAC & NSL-KDD & 84.15 & 84.27 & 84.15 & 83.97 \\
        \cite{ren2023mafsids} & 2023 & MAFSIDS & NSL-KDD & 99.10 & / & / & 99.10 \\
        \cite{benaddi2022robust} & 2022 & Deep SARSA & NSL-KDD & 84.36 & 84.71 & 84.36 & 84.40 \\
        \cite{benaddi2022robust} & 2022 & DQN & NSL-KDD & \textbf{99.36} & \textbf{99.07} & \textbf{99.36} & \textbf{99.21} \\
        \cite{wang2022deep} & 2022 & Dueling DQN & NSL-KDD & 80.31 & 79.62 & 59.87 & 62.62 \\
        \cite{lopez2021network} & 2021 & DRL+RBFNN & NSL-KDD & 90.70 & 87.30 & 96.40 & 92.30 \\
        \cite{dong2021network} & 2021 & SSDDQN & NSL-KDD & 79.43 & 82.81 & 79.43 & 76.22 \\
        \cite{sethi2021attention} & 2021 & A-DQN & NSL-KDD & 97.20 & 96.50 & 99.10 & 97.80 \\
        \cite{ma2020aesmote} & 2021 & DQN & NSL-KDD & 82.09 & 84.11 & 82.09 & 82.43 \\
        \cite{sethi2020robust} & 2020 & DDQN & NSL-KDD & 83.4 & / & / & / \\
        \cite{liu2021reinforcement} & 2020 & DQN & NSL-KDD & 98.71 & 97.35 & 98.71 & 98.30 \\
        \cite{hsu2020deep} & 2020 & DQN & NSL-KDD & 91.40 & 92.80 & 90.20 & 91.48 \\
        \cite{sethi2020context} & 2019 & DQN & NSL-KDD & 81.80 & / & / & / \\
        \bottomrule
    \end{tabular*}
    \label{NSL-KDD-performances}
\end{table*}

\begin{table*}[ht]
    \centering
    \caption{Deep reinforcement learning model performances on other public datasets}
    \begin{tabular*}{\textwidth}{@{\extracolsep{\fill}} llllcclllll}
        \toprule
        \textbf{Reference} & \textbf{Year} & \textbf{Method} & \textbf{Dataset} & \multicolumn{4}{c}{\textbf{Best Model Performance}} \\
        \cmidrule(lr){5-8}
        & & & & \textbf{ACC(\%)} & \textbf{PR(\%)} & \textbf{RC(\%)} & \textbf{F1(\%)} \\
        \midrule
        \cite{mouyart2023multi} & 2023 & AE-DQN & CMU-CERT & 88.80 & 89.10 & 90.70 & 89.90 \\
        \cite{benaddi2022robust} & 2022 & Deep SARSA & UNSW-NB15 & 85.09 & / & / & / \\
        \cite{lopez2021network} & 2021 & DRL+RBFNN & UNSW-NB15 & 82.62 & 82.40 & 82.60 & 82.49 \\
        \cite{hsu2020deep} & 2020 & DQN & UNSW-NB15 & \textbf{91.80} & \textbf{93.20} & \textbf{91.70} & \textbf{92.44} \\
        \cite{sethi2020context} & 2019 & DQN & UNSW-NB15 & / & 68.26 & 86.19 & 76.17 \\
        \cite{li2023soft} & 2023 & AE-SAC & CIC-IDS2017 & 96.65 & 89.10 & 90.70 & 89.90 \\
        \cite{lopez2021network} & 2021 & DRL+RBFNN & CIC-IDS2017 & \textbf{99.70} & \textbf{99.60} & \textbf{99.70} & \textbf{99.60} \\
        \cite{sethi2021attention} & 2021 & A-DQN & CIC-IDS2017 & 98.70 & 98.60 & 99.40 & 98.90 \\
        \cite{ren2023mafsids} & 2023 & MAFSIDS & CIC-IDS2018 & 96.18 & / & / & / \\
        \cite{ren2022id} & 2022 & ID‐RDRL & CIC-IDS2018 & \textbf{96.80} & \textbf{100.00} & \textbf{94.33} & \textbf{96.30} \\
        \cite{ren2022id} & 2022 & ID‐RDRL & CIC-IDS2018 & 96.20 & / & / & 94.90 \\
        \cite{pashaei2023honeypot} & 2023 & ADQN & CIC-IDS2019 & \textbf{99.60} & \textbf{99.30} & \textbf{99.60} & \textbf{99.40} \\
        \cite{ren2022id} & 2022 & DQN+CNN & CIC-IDS2019 & 97.69 & 98.10 & 96.65 & 97.14 \\
        \cite{ren2022id} & 2021 & DRL+RBFNN & CIC-IDS2019 & 99.00 & / & / & / \\
        \cite{li2023soft} & 2023 & AE-SAC & AWID & \textbf{98.98} & \textbf{98.96} & \textbf{98.98} & \textbf{98.92} \\
        \cite{lopez2021network} & 2021 & DRL+RBFNN & AWID & 95.50 & 91.40 & 95.50 & 93.40 \\
        \cite{dong2021network} & 2021 & SSDDQN & AWID & 98.19 & 98.40 & 98.19 & 98.22 \\
        \cite{sethi2020context} & 2019 & DQN & AWID & 96.12 & / & / & / \\

        \bottomrule
    \end{tabular*}
    \label{other-performances}
\end{table*}

DRL-based intrusion models have reached many interesting results, some researchers claim they as a state-of-the-art methods among all intrusion detection models. In this section, we are going to introduce the model performances of DRL-based intrusion detection models in different datasets. \Cref{NSL-KDD-performances} shows the performance of DRL models on the NSL-KDD dataset in recent years. Models such as Big-IDS, MAFSIDS, and A-DQN demonstrate excellent performance 
in terms of accuracy and F1 score. Notably, the DQN model proposed by \cite{benaddi2022robust} achieves the highest accuracy of 99.36\% and F1 score of 99.21\% on NSL-KDD dataset, while MAFSIDS also shows strong performance with 99.10\% accuracy. It is worth noting that multiple DQN-based models in \Cref{NSL-KDD-performances} exhibit significant performance variations (ranging from 81.80\% to 99.36\%), which can be attributed to differences in network architectures, feature engineering 
methods, and hyperparameter configurations employed by different researchers. Different models also excel in precision and recall, with the DRL+RBFNN model showing a balanced performance across these metrics. \Cref{other-performances} lists the performances of DRL models on the UNSW-NB15 and CMU-CERT datasets. Overall, these models perform well on the UNSW-NB15 dataset, with the DQN model achieving an accuracy of 91.80\% and an F1 score of 92.44\%. In comparison, the AE-DQN model also performs notably well on the CMU-CERT dataset, with an accuracy of 88.80\% and an F1 score of 89.90\%. \Cref{other-performances} shows the performance on the CIC-IDS2017 dataset, models like DRL+RBFNN and A-DQN perform excellently across all metrics, with the DRL+RBFNN model achieving an accuracy of 99.70\%, and precision and F1 scores of 99.60\% and 99.60\%, respectively. \Cref{other-performances} also presents the performance on the CIC-IDS2018 dataset. The ID-RDRL model stands out, particularly in recall and F1 score, achieving 100.00\% and 96.30\%, respectively. The performance on the CIC-IDS2019 dataset is also shown in \Cref{other-performances}. The ADQN model excels in all metrics, especially in accuracy (99.60\%) and F1 score (99.40\%). The DQN+CNN model also performs well in terms of precision and recall. For the performance on the AWID dataset, models like AE-SAC and SSDDQN show outstanding performance across all metrics, particularly the AE-SAC model, with an accuracy of 98.98\%, and precision and F1 scores of 98.96\% and 98.92\%, respectively.

\subsection{Comparison with Non-DRL Baseline Methods}
An important limitation of this survey is that \Cref{NSL-KDD-performances} and \Cref{other-performances} exclusively present DRL-based methods without including performance comparisons with non-DRL baseline approaches, which may inadvertently overemphasize the advantages of DRL in network intrusion detection. Recent studies have demonstrated that traditional machine learning algorithms such as XGBoost, Random Forest, and SVM can achieve comparable or even superior performance on benchmark datasets\cite{disha2022performance,waghmode2025intrusion}. Specifically, on the NSL-KDD dataset, Random Forest has achieved accuracies ranging from 99.28\% to 99.88\%, while XGBoost has reached 99.91-99.97\% accuracy\cite{sowenhancing,negandhi2019intrusion}. And SVM combined with Naïve Bayes has attained 99.35\% accuracy\cite{disha2022performance}. These results are comparable to the best-performing DRL methods reported in Table III (e.g., 99.36\% accuracy for DQN). Furthermore, NSL-KDD is increasingly recognized as a relatively simple and outdated dataset that may not adequately represent modern cyber-attacks \cite{disha2022performance}, suggesting that achieving high accuracy on this dataset alone does not necessarily indicate superiority of one approach over another. Future research should incorporate comprehensive comparisons with state-of-the-art non-DRL methods across multiple contemporary datasets to more objectively evaluate the practical advantages of DRL-based approaches in network intrusion detection, particularly in terms of adaptability, real-time learning capabilities, and performance on more complex, realistic datasets that better reflect current threat landscapes.

\subsection{Improved network feature engineering}
Typically, the dataset of network traffic comprises various features that reflect the status of network traffic. However, not all of these features are beneficial for constructing DRL-based network intrusion detection systems. The accurate representation of traffic status using network traffic features is crucial. Consequently, many research efforts have been proposed to effectively extract network features that can accurately represent the actual status of network traffic. Liu \emph{et al.} \cite{liu2020cpss} proposed a method that combines Local-Sensitive Hashing (LSH) with Deep Convolutional Neural Networks (DCNN), selecting optimal features by assessing the distribution of information entropy across each feature value. Ren \emph{et al.} \cite{ren2022id} suggest combining the Recursive Feature Elimination (RFE) and decision tree to select the optimal sub-feature set and the method effectively identifies and eliminates approximately 80\% of the redundant features from the original dataset. Ren \emph{et al.} \cite{ren2023mafsids} further employed Graph Convolutional Networks (GCN) to extract deep features from network data. They transformed the selected input data into dynamic graph networks. Through the hierarchical structure of GCN, more rich and abstract features were extracted. Finally, they combined a multi-agent learning framework to transform the traditional feature selection space of $2^N$ into a competition among $N$ feature agents, effectively reducing the feature space.

\subsection{Handling unbalanced datasets}
The volume of normal network traffic data significantly exceeds that of intrusion data, which is a common-sense observation \cite{gupta2021lio}. Consequently, most network intrusion detection datasets suffer from a severe imbalance in attack-type distribution \cite{abdelkhalek2023addressing}. Therefore, the challenge of training effective DRL-based NID models on an imbalanced dataset have consistently attracted attention. Researchers have attempted to propose various methods to address such issues. Lopez \emph{et al.} \cite{lopez2021network} addressed this problem in NID by augmenting Radial Basis Function (RBF) \cite{buhmann2000radial} neural networks and integrating them with offline reinforcement learning algorithms. They validated the superior performance of this approach across five commonly used datasets. However, the proposed method may face challenges with larger action spaces and have higher computational costs in training. Mohamed \emph{et al.} \cite{mohamed2023deep} utilized a deep State-Action-Reward-State-Action (SARSA) algorithm \cite{rummery1994line} combined with Deep Neural Networks (DNN) \cite{bengio2009learning} to address the issue of data imbalance in NIDs. Although the performance was outstanding, the authors did not analyze the potential drawbacks of their proposed SARSA algorithm. Caution should be exercised when applying this approach. Pashaei \emph{et al.} \cite{pashaei2023honeypot} introduced an adversarial DRL model combined with intelligent environment simulation, presenting an effective approach to addressing the issue of high-dimensional data imbalance in NIDs. This method improves overall classification performance by increasing attention to minority classes, particularly demonstrating its efficiency and effectiveness in practical applications within IoT environments. 

\subsection{Training efficiency}
Normally, network traffic data exhibit high levels of uncertainty and complexity \cite{liu2022complexity}, leading to low training efficiency of DRL, which has been a long-standing concern \cite{nakhaei2025improving}. Louati \emph{et al.} \cite{louati2024big} developed a distributed multi-agent reinforcement learning approach for distributed intrusion detection in large-scale network environments, termed Big-IDS. While the model demonstrated impressive performance, a notable drawback is its low training efficiency. Training the model takes approximately three days when encryption is used and about 12 hours without encryption. This significant time requirement highlights the need for optimization in training procedures to enhance practical applicability. Li \emph{et al.} \cite{li2023soft} introduced a network intrusion detection model named AE-SAC, based on adversarial environment learning and the Soft Actor-Critic (SAC) DRL algorithm. While AE-SAC achieved excellent performance in terms of accuracy and F1 score, its complex network architecture resulted in extended training time. During each training session, both the environment agent and the classifier agent are required to update at least three networks, contributing to the lengthy training process. Kalinin \emph{et al.} \cite{kalinin2023decentralized} enhanced the training efficiency of deep reinforcement learning models in Internet of Things (IoT) intrusion detection by implementing lightweight neural network architectures and developing various multi-agent system architectures. This approach also demonstrated superior performance in terms of accuracy and completeness metrics.

\subsection{Identifying minority and unknown attacks}
Normal network traffic constitutes the major class of the these mentioned dataset. However, in practical applications, the robustness and generalization ability of NID systems are often of greater concern. Therefore, identifying minority types of attacks and recognizing unknown categories of attacks are crucial in the actual deployment of NID systems. Hsu \emph{et al.} \cite{hsu2020deep}developed a DRL-based model for network intrusion detection, equipped with detection and learning modes. This model can switch flexibly based on network traffic behaviours, enabling self-updating capabilities that enhance its ability to recognize unknown network traffic. The study was also tested in a real network environment, where it demonstrated good performance as well. Malik \emph{et al.} \cite{malik2023network} introduced a distributed multi-agent NID system that combines DRL with attention mechanisms \cite{vaswani2017attention}. Utilizing Distributed Q-networks (DQNs) deployed across multiple network nodes, this system offers varied perspectives on the network's security status. It features a multi-agent attention mechanism that increasingly focuses on specific network nodes as performance improves. Additionally, the design incorporates zero-day defence measures to mitigate attacks exploiting unknown vulnerabilities. Notably, the integration of denoising autoencoders (DAE) has significantly enhanced the model's performance. Liu \emph{et al.} \cite{liu2021reinforcement} enhanced a DRL framework by incorporating human operator interaction feedback into the MDP, creating a hybrid structure of Q-networks. They integrated Long Short-Term Memory (LSTM) \cite{graves2012long} networks to better handle time-series features and incorporated a prioritized experience replay mechanism. This innovative approach significantly improved the recognition capabilities for minority and unknown category attacks. Another rising issue is the trade-off between identifying minority attacks and unknown attacks. Dong \emph{et al.} \cite{dong2021network} have reported that when they enhancing the capability to recognize unknown attacks, they have unfortunately seen a decline in the ability to identify specific minority attack types. Xiangyu \emph{et al.} \cite{ma2020aesmote} developed a novel method that combines an enhanced version of the Synthetic Minority Over-sampling Technique (SMOTE) with adversarial RL to improve the detection accuracy of minority classes, such as anomalies or types of attacks. By using SMOTE to generate synthetic samples, they increased the representation of minority classes, thereby addressing the issue of dataset imbalance. Furthermore, they employed two agents within the RL framework, a classifier agent and an environment agent, to facilitate dynamic training and data selection, further optimizing the model's ability to recognize minority classes.


\section{Discussion and future research directions}
\subsection{Datasets}
Although many datasets are currently utilized for training and evaluating NID models, these inherently present several challenges. A primary concern is the imbalance of samples. The collection and tracking of network attacks in the real-world are inherently difficult, leading to a predominance of normal traffic samples over abnormal ones in most datasets. Despite many studies adopting methods such as resampling to balance the training sets \cite{li2023soft, pashaei2023honeypot, malik2023network}, biases and prejudices in the training process are almost inevitable, and there remains a significant gap in methodologies for verifying and evaluating the biases introduced by these training data. It is a common understanding that just because biases are not visible, it does not mean they do not exist. Effective methods for verification need to be developed. Another issue is that most existing benchmark datasets consist of synthetic data generated by cybersecurity researchers simulating attacks in a controlled environment, and it may not well reflect the intrusion behaviours in real network intrusion scenarios. Only the LITNET dataset represents a real-world network attack dataset, containing actual attack traffic produced by attackers in the real-world. However, current research predominantly focuses on the NSL-KDD dataset, and methods based on DRL utilizing the LITNET dataset are still rare, marking an important area for future research.
\subsection{Models}
The performance of various DRL models varies across different datasets. \Cref{NSL-KDD-performances} and \Cref{other-performances} illustrate that different models may exhibit significant performance differences depending on the dataset, suggesting that each model may have unique adaptability to specific types of data. When selecting a deep reinforcement learning model for network intrusion detection, it is essential to consider the model's accuracy, precision, recall, and F1 score comprehensively to choose the most suitable model for the specific application scenario.

Despite recent initial explorations of DRL in NID, it is noteworthy that the structural composition of most DRL policies are still based on simple MLP. According to a recent survey \cite{agarwal2023transformers}, even the most popular transformer architecture has not been widely introduced in DRL-based NID systems, with only a few models like \cite{malik2023network} employing attention mechanisms and achieving good results. One possible reason is that traditional supervised learning and deep learning approaches continue to receive more attention from the network intrusion detection community \cite{yi2023review, he2023adversarial}. Additionally, current NID datasets intuitively seem more suited for supervised learning methods \cite{talukder2024machine}. Recently, the emergence of a novel architecture called Kolmogorov–Arnold Networks (KAN) \cite{liu2024kan}, inspired by Kolmogorov-Arnold representation theorem \cite{kolmogorov1957representation}, has attracted widespread attention in general AI community. This architecture shifts activation functions traditionally located on MLP neurons to the connections/edges between neurons. In their experiments, it replicated the results of a 300,000-parameter MLP-based neural network model using fewer than 200 parameters. Thus, in the foreseeable future, whether KAN could be used to enhance DRL-based NID systems is an interesting topic, especially since most DRL-based NID systems currently approximate optimal strategies using simple MLP structures. Moreover, DRL-based NID seems to lag behind the latest DRL models or methods. Advanced techniques such as PPO \cite{schulman2017proximal} and TRPO \cite{schulman2015trust} have not been embraced by the NID community. DRL-based NID is still in its early stages with many classic DRL-based algorithms. Lastly, the evaluation of the generalization of DRL-based NID models are limited, and relying solely on training and testing splits from a single dataset appears insufficient. In our survey, only a few studies not only trained models but also conducted experiments in real network scenarios \cite{hsu2020deep}. Real-world deployments and tests are urgently needed for DRL-based NID.

\subsection{Tools and Simulation Environments}
\Cref{useful_RL_tools} introduced various reinforcement learning tools and simulation environments that exhibit significant heterogeneity in their purposes and applicability to network intrusion detection (NID) tasks. A notable example is YAWNING-TITAN, which is primarily designed for Autonomous Cyber Defence rather than intrusion detection, focusing on active defence operations such as node isolation and vulnerability patching—a fundamentally different problem space from the passive traffic monitoring and anomaly detection that characterizes traditional NID. Similarly, several other tools are oriented toward penetration testing and offensive security operations rather than intrusion detection. This heterogeneity presents both challenges and opportunities for researchers. While careful selection of simulation environments is essential to ensure alignment with NID requirements, the diverse ecosystem also enables cross-domain learning, where adversarial strategies from penetration testing environments could inform more robust NID systems, and defensive tactics from autonomous cyber defence could inspire novel active learning approaches. Moving forward, the community would benefit from establishing clearer taxonomies that distinguish tools by their primary objectives—detection, defence, or offence—and developing simulation environments specifically tailored to NID tasks with realistic network traffic generation and evaluation metrics aligned with detection performance.

\subsection{Inverse reinforcement learning for NID}
Traditional DRL is known for its inefficiency in sample utilization \cite{yu2018towards, zhang2021sample}. The advent of inverse reinforcement learning may have potential to expedite policy optimization for agents \cite{damiani2022balancing}, it is also critically important for NID systems that require high level of real-time responsiveness. However, current research indicates that IRL has not been applied within the context of DRL-based NIDS. Only some related attempts like Liu \emph{et al.} \cite{liu2021deep} successfully used Inverse Reinforcement Learning (IRL) to reverse-engineer the reward function by observing the trajectories of a trained DQN controller and used this function to design attack strategies that effectively disrupt the control functions of an Industrial Internet of Things (IIoT) system. One possible future direction could involve treating existing network intrusion datasets as expert demonstration data and employing IRL methodologies to model the underlying reward functions of various attacker behaviours. Subsequently, these specialized reward functions could be utilized to train defender agents. For datasets collected from actual network attacks, IRL can be used to effectively model the behaviours of real attackers and potentially train more robust defender agent. Despite this promising research direction, the inherent computational complexity of IRL \cite{wulfmeier2015maximum} algorithms may need the introduction of specific methods for improvement to actualize the theoretical concepts proposed.

\subsection{DRL for intrusion detection to IoT}
In recent years, there has been an increase in research focusing on the network detection for IoT \cite{may2022intrusion, saheed2022machine}. Haosen \emph{et al.} made an initial brief survey on the DRL-based intrusion detection in IoT\cite{zhang2023deep}. As the number of smart devices increases, there is growing concern about how to prevent network intrusions targeting IoT devices\cite{mishra2021internet}. Current approaches are predominantly based on traditional deep learning and machine learning methods as well\cite{khan2022deep}. There is a notable lack of exploration in IoT-specific DRL-based techniques. In our survey, only \cite{kalinin2023decentralized, heartfield2020self}, \cite{ma2020aesmote} and \cite{liu2021deep} have focused on intrusion detection specifically for IoT devices using DRL/RL in recent years. One possible reason for this is the scarcity of benchmark datasets suitable for IoT and DRL. To date, researchers have only conducted preliminary studies on datasets like AWID \cite{chatzoglou2022pick}, Bot-IoT\cite{pacheco2021adversarial}, and IoTID20 \cite{ullah2020scheme}. Nevertheless, with the explosive growth of IoT devices \cite{soori2023internet}, the demand for reliable intrusion detection technologies continues to rise. Therefore, there should remain an optimistic outlook on the application of DRL techniques in IoT intrusion detection. Additionally, researchers from the broad IoT intrusion detection community should continue to strive towards proposing new datasets specifically tailored for IoT device intrusion detection. Moreover, some surveys have also revealed several challenges when introducing DRL into IoT applications, including the need to handle continuous state-action spaces, learning in partially observable environments, ensuring robustness against adversarial attacks, real-time decision making and data privacy and security \cite{chen2021deep, uprety2020reinforcement}.

\subsection{Generative deep model enhanced DRL for NID}
Generative artificial intelligence, such as large language models (LLMs) \cite{chang2023survey}, has recently garnered significant attention worldwide. While some researchers have noted the transformative potential of these methods in the broad field of cybersecurity, few studies have explored how generative AI can be integrated with DRL to advance research in NID. Some researchers have already recognized that LLMs can enhance DRL models in areas such as reward design and world model simulator \cite{cao2024survey}. Consequently, in the foreseeable future, employing LLMs and other generative AI models to augment RL approaches is likely to become a new research paradigm. Therefore, exploring how this potential research paradigm could be extended to NID may become an intriguing research hotspot. Given the powerful capability of LLM-based world models \cite{ge2024worldgpt, yildirim2024task}, one initial idea is to fine-tuning a LLM as a network intrusion expert (attacker)\cite{naito2023llm, fang2024llm} and then using the principle of DRL to train the defender agent through the interaction with LLM-based attacker. Then, using the current public dataset to test and make an evaluation of the defenders. Additionally, the widespread availability of generative AI might also inspire traditional attackers to change their way of attack \cite{fang2024llm}. For instance, with the aid of LLMs, attackers could potentially improve and optimize their attack strategies \cite{fang2024llm}, significantly degrading the performance of NID systems trained on conventional datasets, an urgent concern that should be addressed by the community. Moreover, we cannot guarantee that these potential attackers will cease using these tools as AI technology advances. However, one thing that should be clear is researchers in NID defence systems should responsibly learn to use generative AI to enhance the reliability and defensive capabilities of their proposed systems and make reliable tests as well. Moreover, due to the fact that large simulated datasets exist but fewer real attack examples are available. LLMs attacker might be, to some extent, fitting only to the simulated scenario if the LLMs lack enough knowledge and logic transferring and generalizing ability. It should be a remaining concern in the coming future anyway.

\subsection{Policy functions and architectures}
The findings over recent years suggest that deep neural networks can indeed represent complex policy functions within DRL \cite{li2023deep}. Deep learning has made significant advancements, greatly enhancing agents' capabilities in learning and generalizing from complex environments, which is particularly relevant to network security. Previous surveys also identified a growing interest in integrating attention mechanisms into policy function representations. The promise of attention mechanisms lies in their ability to handle dynamic and high-dimensional data, which is a typical feature in network traffic analysis. However, it is important to note that research on deep model architectures is progressing rapidly. For example, architectures like Mamba \cite{gu2023mamba} and Learn at Test Time (TTT) \cite{sun2024learning} structures have gained popularity recently due to their innovative design and superior performance metrics. These architectures also offer increased capabilities in model interpretability, scalability, and robustness, which are critical for developing effective intrusion detection systems. The need to address this through advanced policy function representations should be a high priority for the network intrusion detection community. By embedding these state-of-the-art deep learning architectures into IDS, researchers may overcome the limitations of conventional methods, offering more precise and efficient threat identification. Implementing such architectures could result in a new generation of IDS that are far better equipped to protect against today's sophisticated cyber threats in our increasingly networked and complex digital environment.

\subsection{Risks of autonomous AI agents in network intrusion}
It is worth noting that with the rapid advancement of autonomous AI agent
technologies, the risk of their malicious exploitation for automated network
intrusion attacks is becoming increasingly prominent. On one hand, autonomous
agents trained via deep reinforcement learning have already demonstrated the
capability to automatically probe vulnerabilities and plan attack paths in
penetration testing scenarios~\cite{hu2021automated}. On the other hand,
LLM-based open-source autonomous agents, exemplified by
OpenClaw~\cite{openclaw2025}, have recently attracted widespread attention.
Such agents are capable of autonomously executing system commands, browser
automation, and file management operations, with their open-source nature and
extensible plugin ecosystems providing remarkable functional flexibility.
However, security researchers have already identified critical vulnerabilities
in such agent frameworks, including remote code execution (e.g.,
CVE-2026-25253~\cite{cve2026openclaw}) and data exfiltration and prompt
injection risks within the skill supply chain. When such autonomous
agents---whether driven by deep reinforcement learning or large language
models---are deliberately leveraged by adversaries, they could autonomously
identify target network vulnerabilities, formulate attack strategies, and
execute intrusion operations without human intervention, posing a serious
challenge to existing network intrusion detection
systems~\cite{fang2024llm}. Therefore, while the cybersecurity community
actively harnesses AI technologies to enhance defensive capabilities, it must
also fully recognize the potential risks of autonomous agent technologies being
misused, and devote efforts to developing detection and defence mechanisms
specifically targeting autonomous agent-driven attacks.

\section{Conclusion}
Network intrusion detection has attracted much attention in broader cybersecurity in recent years. This paper presents a brief survey focused on the application of DRL techniques in the domain of network intrusion detection over the past few years. Current DRL-based network intrusion detection systems are still facing many challenges, including long time for model training, low training efficiency, and a lack of real-world deployment and evaluation. Furthermore, DRL-based NID seems to lag behind the developments in the mainstream DRL community, yet it holds significant potential for growth. Many novel DRL algorithms have not been extended to the application in NID. Additionally, we advocate for the use of data from real-world attack scenarios to train DRL-based NID systems and call for preliminary explorations into modelling attackers' behaviours using inverse reinforcement learning and also advocate for researchers focusing on the intrusion detection for internet of things by collecting more suitable datasets. Lastly, we discussed how the world models built on large language models, or even broader large models can be used to train defensive agents, further facilitating the development of generative-augmented DRL for advancing DRL in the field of NID.

\section*{Ethical Statement}
This research is a survey study that reviews and analyzes existing published literature. It does not involve any human participants, animal subjects, or the collection of personal data. Therefore, no ethical approval was required for this work.

\section*{Data Availability Statement}
This article is a survey paper and does not generate any new datasets. All datasets discussed in this work are publicly available benchmark datasets, and their sources and access links have been cited and referenced within the manuscript accordingly.

\bibliography{paper}

\end{document}